\documentstyle[aps,prl,floats,graphicx,epsfig]{revtex}
\begin{document}
\draft
\twocolumn[\hsize\textwidth\columnwidth\hsize\csname @twocolumnfalse\endcsname

%\documentstyle[preprint,eqsecnum,aps]{revtex}
%\documentstyle[eqsecnum,aps]{revtex}
%\begin{document}
%\draft
%\twocolumn[\hsize\textwidth\columnwidth\hsize\csname @twocolumnfalse\endcsname
%
%\preprint{Preprint }
\title{ Axial anomaly in $^3$He-A: Simulation of Baryogenesis and
Generation of primordial magnetic fields in Manchester and Helsinki}
\author{G.E. Volovik  }

\address{  Low Temperature Laboratory, Helsinki
University of Technology, Box 2200, FIN-02015 HUT, Finland\\
and\\ Landau Institute for Theoretical Physics, Moscow, Russia}

\date{\today} \maketitle

\
\begin{abstract}
The gapless fermionic excitations in superfluid $^3$He-A have a
 ``relativistic''
spectrum close to the gap nodes.
They are the counterpart of chiral particles (left-handed and right-handed)
in high energy physics above the electroweak transition. We discuss the
effective gravity and effective gauge fields induced by these massless
fermions in the low-energy corner. The interaction of the chiral fermions with
the gauge field in $^3$He-A is discussed in detail. It gives rise to the effect
of axial anomaly: Conversion of charge from the coherent motion of the
condensate (vacuum) to the quasiparticles (matter). The charge of the
quasiparticles is thus not conserved. In other words, matter can be created
without creating antimatter. This effect is instrumental for vortex
dynamics, in which the vortex is the mediator of conversion of linear
momentum from the condensate to the normal component via spectral flow in
the vortex core. The same effect leads to the instability of the counterflow in
$^3$He-A, in which the flow of the normal component (incoherent degrees of
freedom) is transformed to the order parameter texture  (coherent degrees of
freedom). We discuss the analogues
of these phenomena in high energy physics. The conversion of the momentum from
the vortex to the heat bath is equivalent to the nonconservation  of baryon
number in the presence of textures and cosmic strings. The counterflow
instability is equivalent to the generation of the hypermagnetic field via the
axial anomaly.  We discuss also an analogue of axions and different
sources of the mass of the ``hyperphoton'' in $^3$He-A.
\end{abstract}
%\eject

\
]
%\narrowtext
%\twocolumn

\section{Introduction}

\subsection{Effective electrodynamics and gravity in $^3$He-A.}

\begin{figure}[!!!t]
%\centerline{\epsfxsize=0.40\textwidth\epsfbox{Lammi1.eps}}
%\bigskip
\begin{center}
\leavevmode
\epsfig{file=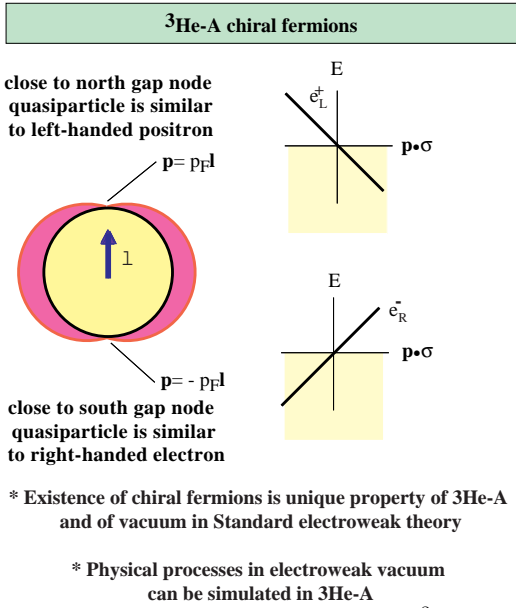,width=0.8\linewidth}
\caption[ChiralFermions]
   { Gap nodes and chiral fermions in $^3$He-A. Close to the gap nodes
the gapless quasiparticles correspond to massless fermions moving
in electromagnetic and gravity fields and obey relativistic
equations. The left-handed particles have positive charge
$e=+1$ and are in the vicinity of the north pole, as determined by the
direction
of the ${\hat {\bf l}}$-vector, while the right particles with
$e=-1$ are in the vicinity of the south pole. If the  ${\hat{\bf
l}}$-field is not uniform in space and time, the space and time derivatives of
${\bf A}=p_F{\hat{\bf l}}$ act on the quasiparticle similar to magnetic and
electric fields.}
\label{ChiralFermions}
\end{center}
\end{figure}

Many aspects of high energy physics can be modelled in condensed matter
\cite{Wilczek}.  Superfluid $^3$He-A provides a
rich source for such a modelling.
The most pronounced property of this superfluid is
that in addition to the numerous bosonic fields (collective modes of the
order parameter) it contains gapless fermionic quasiparticles. Close to the
gap nodes, the points in  momentum space where the energy is zero
(Fig.~\ref{ChiralFermions}), the energy spectrum of  quasiparticles is
linear in
momentum ${\bf p}$. This simple circumstance has far-reaching
consequences:
The
low energy fermions and some of the bosons obey ``relativistic''
 equations, while
their interaction with the superfluid vacuum
mimics that of elementary particles with gauge
fields.  This illustrates the principle \cite{Nielsen} that the effective
physics in a low energy corner becomes more symmetric than in the general case.
In $^3$He-A we
have two low energy corners, at ${\bf p}\approx \pm p_F{\hat{\bf l}}$, where
$p_F$ is the Fermi momentum and ${\hat{\bf l}}$ the unit vector specifying
the direction of the nodes. This picture does not depend on details of the
underlying microscopic interactions of atoms, whose only role is to
produce
values of ``fundamental constants'', such as the ``speed of light'' and
the ``Planck energy''.

Close to the gap node the square of the quasiparticle energy $E$ is generally
a quadratic form of the deviation of the momentum ${\bf p}$ from the position
of the nodes $ \pm p_F{\hat{\bf l}}$:
\begin{equation}
E_\pm^2({\bf p})=g^{ik}(p_i \mp p_F{\hat  l}_i) (p_k \mp p_F{\hat  l}_k)~.
\label{E^2form}
\end{equation}
Let us introduce an effective vector potential of the ``electromagnetic field''
\begin{equation}
{\bf A}= p_F{\hat{\bf l}}~,
\label{vector potential}
\end{equation}
and the ``electric charge'' $e$, with  $e=+ 1$ for the
quasiparticles in the vicinity of the node at
$p_F{\hat{\bf l}}$ and $e=-1$  for the
quasiparticles in the vicinity of the opposite node, at
$-p_F{\hat{\bf l}}$. Then
one obtains a
spectrum of relativistic fermions moving on the gravitational and
electromagnetic background, determined by the metric tensor $g^{ik}$ and the
vector potential ${\bf A}$:
\begin{equation}
E^2({\bf p})=g^{ik}(p_i -eA_i) (p_k-eA_k)~.
\label{E^2relativistic}
\end{equation}

The symmetric matrix $g^{ik}$, which gives the contravariant components of
the metric
tensor, is generally determined by the directions of
the principal axes forming the orthonormal basis (${\hat{\bf e}}_1,{\hat{\bf
e}}_2,{\hat{\bf e}}_3$) and by the ``speeds of light'' along these directions:
\begin{equation}
g^{ik}=c_1^2 \hat e_1^i \hat e_1^k+c_2^2 \hat e_2^i \hat e_2^k+c_3^2 \hat e_3^i
\hat e_3^k~.
\label{3speeds}
\end{equation}

It is important that the effective gauge field ${\bf A}$ and the effective
metric $g^{ik}$ depend on space and time, since the order parameter
in general and the ${\hat{\bf l}}$-vector in particular are not fixed in
$^3$He-A and can form different types of textures. The quasiparticles view
the order parameter textures as a curved Lorentzian space-time and
{\em simultaneously} as gauge fields.
These fields are dynamical: The Effective Lagrangian for
``electromagnetic'' and ``gravitational'' fields can be obtained by integrating
over the fermionic field. The same principle was used by Sakharov and Zeldovich
to obtain an Effective Gravity \cite{Sakharov} and Effective Electrodynamics
\cite{Zeldovich}  from vacuum fluctuations. In some special cases
(and in this review we consider just such a case) the main contribution to the
effective action comes from the vacuum fermions whose momenta ${\bf p}$ are
concentrated near the gap nodes, i.e. from the ``relativistic'' fermions. In
these (and only in these) cases one obtains an effective Lagrangian which
 gives Maxwell equations for the ${\bf A}$-field.  Since the
``photons'' are thus constructed from the fermionic degrees of freedom, the
metric $g^{ik}$, which governs the propagation of ``photons'', is the same
as the
metric governing the dynamics of the underlying fermionic
quasiparticles.
% This means that  the ``light'' also propagates with the ``speed of light''.
Following
the title of the Laughlin talk at this Symposium, this provides
an example of a ``Gauge Theory from Nothing''
\cite{Laughlin}.

From Eq.(\ref{E^2relativistic}) it follows that $g^{00}=-1$ and
$g^{0i}=0$, but this is not the general case: Typically all the
components of the dynamical metric tensor $g_{\mu\nu}$ depend on the
position in
space-time. In some cases the effective metric is not trivial giving rise
to conical singularities \cite{VolovikGravity1997},  event
horizons and ergoregions \cite{JacobsonVolovik,RotatingCore}. This also allows
to simulate quantum gravity. Note that the primary quantities in this
effective (quantum) gravity are the {\em contravariant} components
$g^{\mu\nu}$.
They appear in the low-energy corner of the fermionic spectrum and represent
the low-energy properties of the quantum vacuum. The geometry of the effective
space-time, in which
the free quasiparticles follow a geodesic, is determined by the inverse metric
$g_{\mu\nu}$ and thus is a secondary object.  In a similar manner the
effective Lorentzian space-time comes from the spectrum of the sound
waves propagating on the background of a moving inhomogeneous liquid
\cite{UnruhSonic,Jacobson1991,Visser1997}. The difference to the case of
superfluid liquid $^3$He-A is that ordinary liquids are essentially
dissipative classical systems and thus cannot serve as a model of the quantum
vacuum.

The above mechanism of the generation of the gauge field {\bf A}
and gravity $g^{\mu\nu}$ is valid
for a general system with point gap nodes. In the particular case of
$^3$He-A the initial ``nonrelativistic'' fermionic spectrum has the form
\begin{equation}
E^2({\bf p})= v_F^2(p-p_F)^2 +{\Delta_0^2\over
p_F^2}({\hat{\bf l}} \times {\bf p})^2,
\label{AphaseSpectrum}
\end{equation}
where $\Delta_0$ is the gap amplitude; $v_F=p_F/m^*$ is the Fermi
velocity and $m^*$
the effective mass of the quasiparticle in the normal Fermi-liquid
state, which is typically about 3--6 times the bare mass $m_3$  of the
$^3$He atom.

In the low energy corner one obtains the  ``relativistic''
spectrum of  Eq. (\ref{E^2relativistic}) with the following values of the
``fundamental constants'' \cite{exotic,VolovikVachaspati}
\begin{equation}
c_1=c_2={\Delta_0\over
p_F}\equiv c_{\perp} ~,~c_3=v_F\equiv c_{\parallel}~.
\label{SpeedsInAphase}
\end{equation}
The space characterizing the motion of quasiparticles in $^3$He-A, i.e. the
space in which the quasiparticles move along the geodesic curves (in the
absence of other forces) has an uniaxial anisotropy, with the anisotropy axis
along
${\hat{\bf l}}$:
\begin{equation}
{\hat{\bf e}}_3={\hat{\bf l}}~.
\label{e3}
\end{equation}
The speed of a ``light'' along the ${\hat{\bf l}}$-vector,
$c_{\parallel}$, is about 3 orders of magnitude larger than that in the
transverse direction:
$ c_{\parallel}\gg c_{\perp}$.
Another important fundamental constant, $\Delta_0$, plays the role of the
Planck energy cut-off, as will be illustrated later on.

\subsection{Chiral fermions in $^3$He-A}

The chiral  properties of the  fermionic spectrum is revealed after the
square root of Eq.~(\ref{E^2relativistic}) is taken. This procedure is not
unambiguous: One has to use the underlying BCS theory of Cooper pairing,
which leads to the superfluid A-phase state in
$^3$He. In BCS theory one obtains the Bogoliubov-Nambu
Hamiltonian for fermions, which in the low-energy corner transforms into the
Weyl Hamiltonian for massles chiral particles. It is represented by the  proper
square root of Eq.~(\ref{E^2relativistic}):
\begin{equation}
{\cal H}=-e\sum_a c_a \tau^a \hat e^i_a(p_i- eA_i) ~,
\label{WeylHamiltonian}
\end{equation}
where $\tau^a$ are the Pauli matrices acting in the Bogolibov-Nambu
particle-hole space.

\begin{figure}[!!!t]
%\centerline{\epsfxsize=0.40\textwidth\epsfbox{LammiNew.eps}}
%\bigskip
\begin{center}
\leavevmode
\epsfig{file=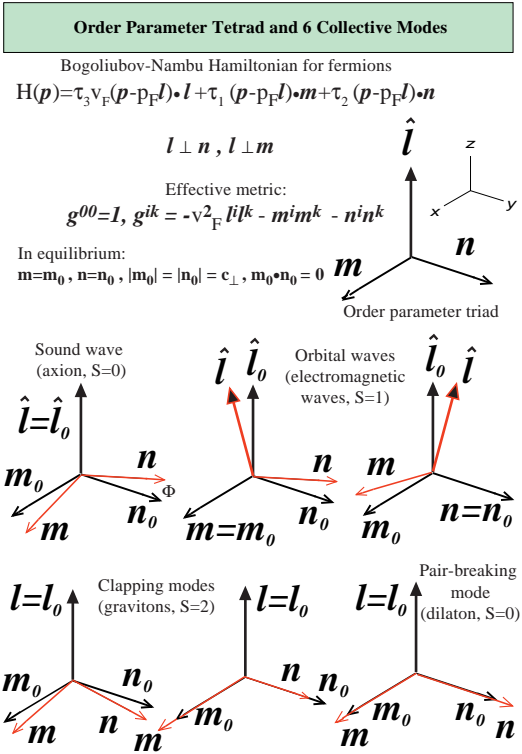,width=0.8\linewidth}
\caption[CollectiveModes]
   {Triads ${\bf m}$, ${\bf n}$ and $\hat l \propto {\bf m}\times {\bf n}$,
which characterize the order parameter in $^3$He-A for a given spin projection.
The dynamics of two vector fields ${\bf m}$ and ${\bf n}$ gives rise to 6
collective modes: (1) Sound wave (axion) is propagating rotational oscillations
of the triad around the $\hat l$ vector. The angle $\Phi$ of this rotation is
the phase of the superfluid condensate, whose gradient determines the
superfluid velocity; (2-3)  Orbital waves (photons) are   propagating
oscillations of the direction of the
$\hat l$ vector; (4-5) Clapping modes (gravitons) are out-of-phase oscillations
of the diad ${\bf m}$,
${\bf n}$; and (6) Pair-breaking mode (dilaton) is the oscillation of the
modulus of the
${\bf m}$ and ${\bf n}$ vectors. This corresponds to the propagating
oscillations of the transverse speed of light $c_\perp$. If the spin degrees of
freedom are taken into account, the number of collective modes is
triplicated.}
\label{CollectiveModes}
\end{center}
\end{figure}

The more close inspection of the BCS theory for $^3$He-A reveals that the order
parameter contains 18 degrees of freedom and thus 18 propagating collective
modes. Six of these collectives modes, which represent propagating oscillations
of position of nodes and of the slopes of the energy spectrum at the nodes, are
shown in  Fig.~\ref{CollectiveModes} together with  their analogs
in relativistic theories.

The important property of this Hamiltonian is that the sign of the ``electric''
charge $e$ simultaneously determines the chirality of the fermions. This is
clearly seen with a simple isotropic example having $c_1=c_2=c_3=c$:
\begin{equation}
{\cal H}=-e c\vec\tau\cdot({\bf p}- e{\bf A}) ~.
\label{WeylIsotropic}
\end{equation}
A particle  with positive (negative) $e$ is left-handed (right-handed): Its
Bogoliubov spin $\vec \tau$ is antiparallel (parallel) to the momentum ${\bf
p}$, if ${\cal H}$ is positive definite.
Thus the field ${\bf A}$ corresponds to the axial field in relativistic
theories. The  symmetry between left and right is broken in $^3$He-A.

%%%%%%%%%%%%%%%%%%%%%%%%%%%%%%%%%%%%%%%%%%%%%%%%%%%%%%%%%%
%%%%%%%%%%%%%%%%%%%%%%%%%%%%%%%%%%%%%%%%%%%%%%%%%%%%%%%%%%

Rather few systems have (3+1)-dimensional chiral fermions as excitations. The
superfluid
$^3$He-A and the Standard Model of the electroweak interactions are among these
exotic systems. This is why
$^3$He-A is the best condensed matter system for the simulation of effects
related to the chiral nature of the fermions, especially of the  chiral
anomaly. There are  other condensed matter systems with chiral fermions, but
these fermions occupy a space-time of 2+1 or 1+1 dimensions. Examples are the
(2+1)-dimensional fermions in high-temperature superconductors \cite{SimonLee};
(1+1)-dimensional chiral edge states in the quantum Hall
effect \cite{WenEdgeStates,StoneEdgeStates}, and in superconductors with broken
time-reversal symmetry
\cite{LaughlinT-symmetry,VolovikT-symmetry}. Finally
fermionic excitations in the core of quantized vortices
 bear this property, too \cite{RotatingCore}.
The gap nodes in (3+1)-dimensional theories can appear also in
different types of ``color superfluidity'' -- quark condensates in
dense baryonic matter
\cite{ColorSuperconductivity1,ColorSuperconductivity2}
(The quark condensate phase analogous to the superfluid $^3$He-B, where
color and
flavour are locked together instead of spin and orbital momenta, while the
gap is isotropic and thus has no nodes, was also discussed
\cite{ColorSuperconductivity3}).

\begin{figure}[!!!t]
%\centerline{\epsfxsize=0.40\textwidth\epsfbox{Lammi2.eps}}
%\bigskip
\begin{center}
\leavevmode
\epsfig{file=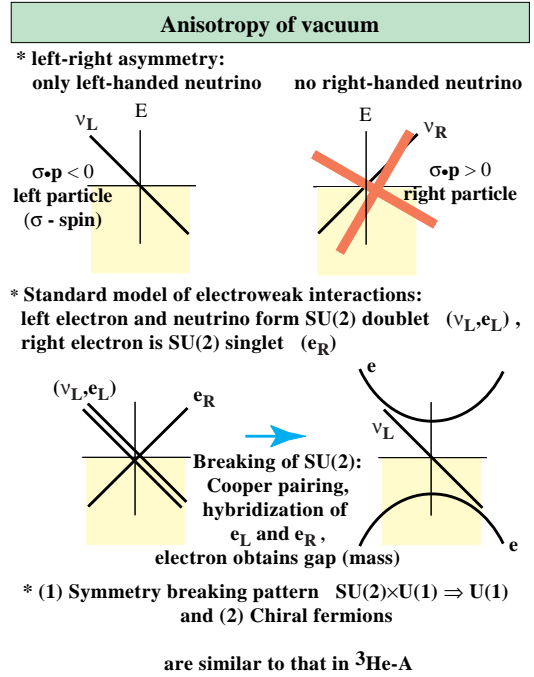,width=0.8\linewidth}
\caption[Metal-Insulator]
    {{\it (top)}: The spectrum of fermionic excitations of the
physical vacuum contains a branch of
chiral particles,  the left-handed neutrino branch. The right-handed
neutrino does not exist (or interacts with other matter only via the
gravitational field).  In the   symmetric state of the early Universe, the
left leptons (neutrino and left electron) formed the
$SU(2)$ doublet, while the right electron was the $SU(2)$ singlet.  {\it
(bottom)}: During the cool down of the Universe the $SU(2)$ symmetry was broken
in a phase transition. The left and right electrons were hybridized forming the
present electronic spectrum with the gap
$\Delta=m_ec^2$. The electric properties of the vacuum thus exhibited  the
metal-insulator transition.}
\label{Metal-Insulator}
\end{center}
\end{figure}

 The spectrum of fermionic excitations of the electroweak
vacuum in the present Universe contains one branch of chiral particles: The
left-handed neutrino branch
(Fig.~\ref{Metal-Insulator}). The right-handed neutrino is not present (or
interacts
 with other matter different from  the left-handed one).  This is a
remarkable manifestation of the violation of the left-right symmetry in the
electroweak vacuum. Another symmetry, which is broken in the present Universe,
is the $SU(2)$ symmetry of weak interactions. In the symmetric state of
the early Universe, the left leptons (neutrino and left electron)
formed a $SU(2)$ doublet, while the right electron is in a $SU(2)$ singlet.
During
the cooldown of the Universe the phase transition occured,
at which the $SU(2)\times U(1)$ symmetry was broken to the
electromagnetic $U(1)$ symmetry. As a consequence, the left and right electrons
were hybridized forming the present electronic spectrum with the gap
$\Delta=m_ec^2$. The electric properties of the vacuum  thus exhibited the
metal-insulator phase transition: The ``metallic'' state of the vacuum with the
Fermi point in the elecronic spectrum was transformed to the insulating state
with the gap.  Recent numerical calculations suggest that this
transformation occurs either by the first order phase transition or by
continuous cross-over without any real symmetry breaking \cite{Kajantie}.

%%%%%%%%%%%%%%%%%%%%%%%%%%%%%%%%%%%%%%%%%%%%%%%%%%%%%%%%%%
%%%%%%%%%%%%%%%%%%%%%%%%%%%%%%%%%%%%%%%%%%%%%%%%%%%%%%%%%%

The similarity between the chiral fermions in electroweak theory and in
$^3$He-A has also a topological origin. The gap nodes -- zeroes in the
particle (quasiparticle) spectrum -- are characterized by a
topological invariant in 4-momentum space
belonging to the third homotopy group $\pi_3$ \cite{exotic}:
\begin{equation}
N_{\rm top} = {1\over{24\pi^2}}e_{\mu\nu\lambda\gamma}~
{\bf tr}\int_{\sigma}~  dS^{\gamma}
~ {\cal G}\partial_{p^\mu} {\cal G}^{-1}
{\cal G}\partial_{p^\nu} {\cal G}^{-1} {\cal G}\partial_{p^\lambda}  {\cal
G}^{-1}~.
\label{TopInvariant}
\end{equation}
Here
\begin{equation}
 {\cal G}(p_\mu)= {1\over ip_0  +{\cal H}}
\label{GreenFunction}
\end{equation}
is the Green's function and $\sigma$ is the 3-dimensional surface around the
point node in the 4-momentum space. For the relativistic chiral particle the
node is at
$p_0=0$, ${\bf p}=0$, while in $^3$He-A the nodes are at $p_0=0$, ${\bf
p}= \pm p_F \hat{\bf l}$. In all cases the topological invariant is nonzero:
$N_{\rm top} =\pm 1$, and the sign of $N_{\rm top}$ depends on chirality.

The topological stability -- the conservation of the topological invariant in
Eq.~(\ref{TopInvariant}) --  is important for the fermionic system. It implies
that under a deformation of the system (under a continuous change of the
system parameters) the gap nodes in momentum space can arise or disappear only
in pairs (node-``antinode'' pairs). This topological stability, which does not
depend on the details and the symmetry of the system, provides topological
conservation of chirality:  The algebraic number of the chiral fermions,
i.e. the number of the right fermionic species minus the number of the left
fermionic species, is conserved:
$\Delta N =  N_{FR} - N_{FL}=\sum N_{\rm top}$. In $^3$He-A one has $  N_{FR}
= N_{FL}=1$ and thus $\Delta N =\sum N_{\rm top}=0$.

In the relativistic theories the electroweak
transition
$SU(2)\times U(1)
\rightarrow U(1)$  satisfies this topological rule: If the right neutrinos are
absent, the algebraic number of  chiral fermions  per
each generation  is $\Delta N = -1$ in both phases: in the symmetric phase
$SU(2)\times U(1)$ one has $\Delta N =7-8= -1$ and  in the broken symmetry
phase $U(1)$ one has
$\Delta N =0-1= -1$. In this case the conservation of the topological
invariant  provides the zero mass for neutrino. In the unification
theories, the
$SU(5)$ symmetry breaking pattern with $N_{FL}=10+5$ left fermions in one
generation does not satisfy this topological rule.  The rule holds only if one
doubles the number of fermions and considers right antiparticles as
independent particles: in this case
$\Delta N =15-15= 0$.  In the $SU(4)\times  SU_L(2)\times SU_R(2)$  theory with
$  N_{FR} = N_{FL}=8$  the topological rule is
satisfied without the doubling of fermions: one has $\Delta N =0$
throughout all the route of the symmetry breaking to $SU(3)\times  U(1)$.

It is important that if the vacuum is characterized by nonzero topological
charge, $\sum N_{\rm top}\neq 0$, the system has massless fermions. This means
that the problem of the neutrino mass is directly related to the momentum space
topology of the vacuum.

%%%%%%%%%%%%%%%%%%%%%%%%%%%%%%%%%%%%%%%%%%%%%%%%%%%%%%%%%%%

%%%%%%%%%%%%%%%%%%%%%%%%%%%%%%%%%%%%%%%%%%%%%%%%%%%%%%%%%%%

\section{AXIAL ANOMALY}

\subsection{Adler-Bell-Jackiw equation}

\begin{figure}[!!!t]
%\centerline{\epsfxsize=0.40\textwidth\epsfbox{Lammi3.eps}}
%\bigskip
\begin{center}
\leavevmode
\epsfig{file=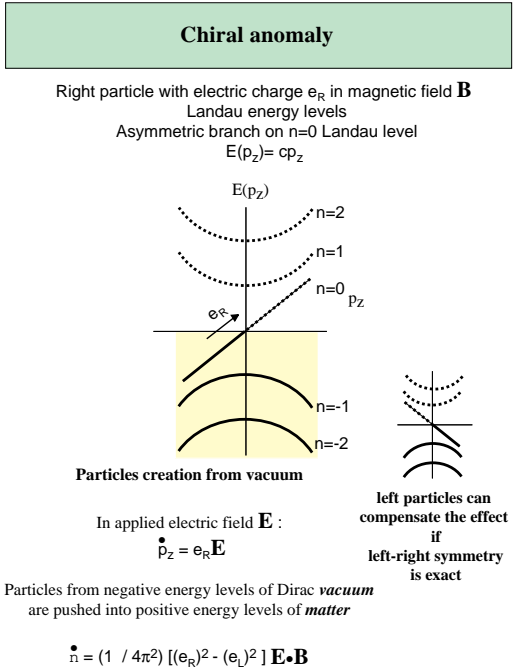,width=0.8\linewidth}
\bigskip
\caption[ChiralAnomaly]
   {Chiral anomaly: the spectral flow from the occupied energy levels caused
by the applied electric field leads to the creation of the fermionic  charge
from the vacuum if the left-right symmetry is violated. }
\label{ChiralAnomaly}
\end{center}
\end{figure}

Chiral fermions interacting with gauge fields exhibit the effect of chiral
anomaly, the nonconservation of matter charge due to the interaction
of matter with the quantum vacuum.
The origin for the axial anomaly can be seen from
the behavior of the chiral particle in a constant magnetic field,
${\bf A}=(1/2){\bf B}\times {\bf r}$.
The Hamiltonians for the right particle with the electric charge $e_R$ and for
the left particle with the electric charge $e_L$ are
\begin{equation}
{\cal H}= c\vec\tau\cdot({\bf p}- e_R{\bf A}) ~, {\cal H}= -c\vec\tau\cdot({\bf
p}- e_L{\bf A}) ~.
\label{WeylForLeftRight}
\end{equation}
Fig.~\ref{ChiralAnomaly} shows the energy spectrum in a magnetic field
${\bf B}$ along $z$; the thick lines show the occupied negative-energy states.
Motion of the particles in the plane perpendicular to ${\bf B}$ is
quantized into the  Landau levels
shown.  The free motion is thus effectively reduced to
one-dimensional motion along ${\bf B}$ with momentum $p_z$.
Because of the chirality of the particles
the lowest ($n=0$) Landau
level is asymmetric. It crosses zero only
in one direction: $E=cp_z$ for the right particle and $E=-cp_z$ for the left
one. If we now apply an electric field ${\bf E}$ along $z$,
particles are pushed from negative to positive energy
levels according to the equation of motion $\dot p_z =e_{R} E_z$ ($\dot p_z
=e_{L} E_z$) and the whole Dirac sea moves up (down)
creating particles and electric charge from the vacuum.
This motion of particles along the
``anomalous'' branch of the spectrum is called
{\em spectral flow}. The rate of particle production is
proportional to  the density of states at the Landau level,
which is  $\propto \vert e_R{\bf B}\vert$ ($\vert e_L{\bf B}\vert)$, so
that the
rate of production of particle number $n=n_R+n_L$ and of charge
$Q=n_Re_R+n_Le_L)$ from the vacuum is
\begin{equation}
\dot{n} ={1\over {4\pi^2}} (e_R^2-e_L^2){\bf E} \cdot {\bf
B} ~,~\dot{Q}={1\over {4\pi^2}} (e_R^3-e_L^3){\bf E} \cdot
{\bf B}  ~.
\label{ChargeParticlProduction}
\end{equation}

%%%%%%%%%%%%%%%%%%%%%%%%%%%%%%%%%%%%%%%%%%%%%%%%%%%%%%%%%
%%%%%%%%%%%%%%%%%%%%%%%%%%%%%%%%%%%%%%%%%%%%%%%%%%%%%%%%%

This is an anomaly equation for the production of particles from
 vacuum of the type found by  Adler\cite{Adler1969} and by Bell and
Jackiw\cite{BellJackiw1969} in the context of neutral
pion decay. We see that for particle or charge
creation from ``nothing'' it is necessary to have an asymmetric branch of the
dispersion relation $E(p)$ which crosses the axis from
negative to  positive energy. Additionally, the symmetry
between the left and right particles has to be violated: $e_R\neq e_L$ for the
charge creation and $e_R^2\neq e_L^2$ for the particle creation.

\subsection{Anomalous nucleation of baryonic charge}

In the standard electroweak model there is an
additional accidental global symmetry $U(1)_B$  whose
classically conserved charge is the baryon number $Q_B$.
Each of the quarks is assigned $Q_B=1/3$  while the leptons (neutrino and
electron) have $Q_B=0$. This baryonic number is not conserved due to the
axial anomaly. There are two gauge fields whose ``electric'' and
``magnetic'' fields
become a source for baryoproduction: The hypercharge field $U(1)$ and the
weak field $SU(2)$. The corresponding hypercharges $Y$ and  weak charges $W$
of the left  $u$ and $d$ quarks are
\begin{equation}
Y_{dL}=Y_{uL}=1/6~,~W_{dL}=-W_{uL}= 1/2    ~,
\label{ChargesOfLeftQuarks}
\end{equation}
whereas for the right $u$ and $d$ quarks one has
\begin{equation}
Y_{uR}= 2/3  ~,~Y_{dR}= -1/3~,~  W_{dR}=W_{uR}= 0  ~  .
\label{ChargesOfRightQuarks}
\end{equation}

Let us first consider the effect of the hypercharge field. Since the number of
different species of quarks carrying the baryonic charge is
$3N_F$  (3 colours
$\times$ $N_F$ generations of fermions) and the baryonic charge of the quark is
$Q_B=1/3$,   the production rate of baryonic
charge in the presence
of hyperelectric and hypermagnetic fields is
\begin{equation}
{ N_F\over 4\pi^2}(Y_{dR}^2 +Y_{uR}^2 - Y_{dL}^2 - Y_{uL}^2)~{\bf B}_Y\cdot
{\bf
E}_{Y}.
\label{BarProdByHypercharge}
\end{equation}
Since the
hypercharges of left and right quarks are different (see
Eqs.~(\ref{ChargesOfLeftQuarks},\ref{ChargesOfRightQuarks})),  one obtains a
nonzero production of baryons by the hypercharge field
\begin{equation}
{N_F\over 8\pi^2}{\bf B}_Y\cdot {\bf E}_{Y}.
\label{BarProdByHypercharge2}
\end{equation}

The weak electric and magnetic
fields also contribute to the production of the baryonic charge:
\begin{equation}
  {N_F\over 4\pi^2} ( W_{dR}^2 +
 W_{uR}^2 -  W_{dL}^2 -  W_{uL}^2)~{\bf B}^a_W\cdot {\bf E}_{aW} ~,
\label{BarProdByWeak}
\end{equation}
which gives
\begin{equation}
 - {N_F\over 8\pi^2}{\bf B}^a_W\cdot {\bf E}_{aW}.
\label{BarProdByHypercharge3}
\end{equation}

Thus the total rate of baryon production in the Standard model takes
the form
\begin{equation}
\dot Q_B= {{N_F} \over {8 \pi^2}} \left (
-  {\bf B}^a_W\cdot {\bf E}_{aW} +
   {\bf B}_Y\cdot {\bf E}_{Y} \right).
\label{TotalBaryoProduction}
\end{equation}
The first term  comes from nonabelian $SU(2)$ fields, it shows that the
nucleation of baryons occurs when the topological charge of the
vacuum changes, say, by sphaleron or due to de-linking of linked loops of
the cosmic strings. The second, nontopological, term describes the
exchange of the baryonic charge between the
hypermagnetic  field and the fermionic degrees of freedom.

\subsection{Anomalous nucleation of linear momentum in $^3$He-A}

The anomaly equation which describes the nucleation of fermionic charges
in the presence of magnetic and electric fields describes both the
production of the baryons  in the electroweak vacuum (baryogenesis) and  the
production of  the  linear momentum in the superfluid
$^3$He-A (momentogenesis).  In $^3$He-A
the effective $U(1)$ gauge field is generated  by the moving
${\hat{\bf l}}$-texture.  According to Eq.~(\ref{vector potential}),  the time
and space dependent
${\hat{\bf l}}$ vector, associated with the motion of the so-called
continuous vortex (see below), produces a force on the excitations
equivalent to
that of an ``electric''  (or ``hyperelectric'') field
${\bf E}=p_F \partial_t {\hat {\bf l}}$ and a ``magnetic''  (or
``hypermagnetic'' field)
${\bf B}=p_F{\bf \nabla}\times {\hat {\bf l}}$ acting on particles of
unit charge.  Equation~(\ref{ChargeParticlProduction})
can then be  applied to calculate the rate at which left-handed
and right-handed quasiparticles are created by spectral flow. What we are
interested in is the production of the particle momentum due to spectral flow:
\begin{equation}
\dot{\bf P}= {1\over {4\pi^2}}({\bf P}_R-{\bf P}_L) ~(
{\bf E} \cdot   {\bf B} \, \, ) ~~.
\label{MomentoProduction1}
\end{equation}
Since the right and left particle
have opposite momenta ${\bf P}_R=p_F{\hat {\bf l}}=-{\bf P}_L$,
excitation momentum is created at a rate
\begin{equation}
\dot{\bf P}= { p_F^3\over {2\pi^2}} \hat {\bf l} ~(
\partial_t \hat {\bf l} \cdot  (\vec\nabla\times \hat {\bf l} )\, \, ) ~~.
\label{MomentoProduction2}
\end{equation}

However, the total linear momentum of the liquid has to be conserved.
Therefore Eq. (\ref{MomentoProduction2}) implies that in the presence of a
time-dependent texture momentum is transferred from the
superfluid ground state (analogue of vacuum) to the heat
bath of excitations forming the normal component (analogue of matter).

\section{SPECTRAL FLOW FORCE on VORTEX}

\subsection{Continuous vortex and baryogenesis in textures}

\begin{figure}[!!!t]
%\centerline{\epsfxsize=0.40\textwidth\epsfbox{Lammi4.eps}}
%\bigskip
\begin{center}
\leavevmode
\epsfig{file=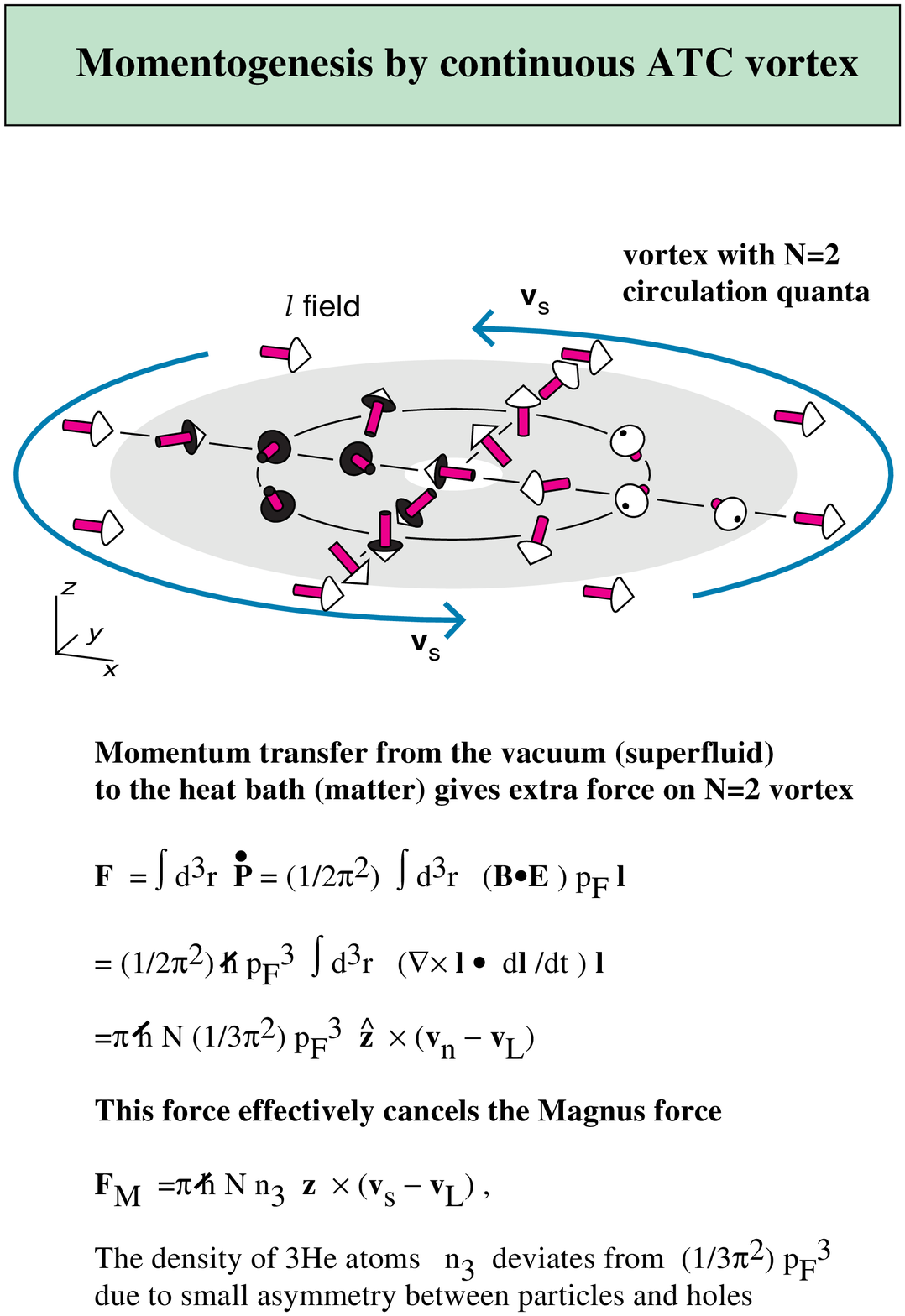,width=0.8\linewidth}
\caption[ContinuousMomentogenesis]
    {The order
parameter ${\hat{\bf l}}$-texture (analogue of the vector potential of the
(hyper) electromagnetic field) in the soft core of the continuous
Anderson-Toulouse-Chechetkin vortex
\cite{Chechetkin,AT} (ATC vortex). The moving vortex converts the textural
(vacuum) fermionic charge (linear momentum) to fermionic
quasiparticles (matter). The moving vortex generates the time dependence of the
order parameter, which is equivalent to an electric field ${\bf E}$.
Together with
the effective magnetic field ${\bf B}$ concentrated in the soft core
of the ATC
vortex this gives rise to a nonzero dot product
${\bf B}\cdot {\bf E}$, which is the anomalous source of fermionic charge. }
\label{ContinuousMomentogenesis}
\end{center}
\end{figure}

The anomalous production of linear momentum leads to
an additional force acting on the continuous vortex in $^3$He-A
(Fig.~\ref{ContinuousMomentogenesis}).

The continuous vortex, first discussed by Chechetkin \cite{Chechetkin} and
Anderson and Toulouse\cite{AT} (ATC vortex), has in its simplest realization
the following distribution of the ${\hat{\bf l}}$-field  ($\hat{\bf z}$,
${\hat{\bf r}}$ and ${\hat{\bf \phi}}$ are unit vectors of the cylindrical
coordinate system)
\begin{equation}
{\hat{\bf l}}(r,\phi)={\hat{\bf z}} \cos\eta(r) + {\hat{\bf
r}} \sin\eta(r)~,
\label{lTextureContVortex}
\end{equation}
where $\eta(r)$ changes from $\eta(0)=\pi$ to $\eta(\infty)=0$ in the so called
soft core of the vortex. The
superfluid velocity
${\bf v}_s$ in superfliud $^3$He-A is determined by the twist of the triad
${\hat{\bf e}}_1,{\hat{\bf e}}_2,{\hat{\bf e}}_3$ and corresponds to
torsion in the tetrad formalism of  gravity  (the space-time dependent
rotation of vectors
${\bf m}=c_\perp {\hat{\bf e}}_1$ and ${\bf n}=c_\perp {\hat{\bf e}}_2$ about
axis $\hat l$ in Fig.~\ref{CollectiveModes}):
\begin{equation}
{\bf v}_s ={\hbar\over 2 m_3} \hat e_1^i\vec \nabla \hat
e_2^i ~.
\label{v_s}
\end{equation}
In comparison to a more familiar singular vortex, the continuous vortex has
a regular
superfluid velocity field
\begin{equation}
{\bf v}_s(r,\phi)= -{\hbar\over 2 m_3 r}[1+\cos\eta(r)]{\hat {\bf \phi}}~~,
\label{v_sContVortex}
\end{equation}
with no singularity on the vortex axis.

%%%%%%%%%%%%%%%%%%%%%%%%%%%%%%%%%%%%%%%%%%%%%%%%%%%%%%%%%%%
%%%%%%%%%%%%%%%%%%%%%%%%%%%%%%%%%%%%%%%%%%%%%%%%%%%%%%%%%%

The stationary vortex generates a
``magnetic'' field. If the vortex moves with a
constant velocity ${\bf v}_L$ it also generates
an ``electric''  field, since ${\hat{\bf l}}$  depends  on
${\bf r}-{\bf v}_Lt$:
\begin{equation}
{\bf B}=p_F\vec \nabla \times {\hat {\bf l}}~~,~~{\bf
E}=\partial_t {\bf A}=-p_F({\bf v}_L\cdot
\vec\nabla){\hat {\bf l}}
\label{EfieldContVortex}
\end{equation}

The net production of the quasiparticle momenta by the spectral flow in the
moving vortex means, if the vortex moves with respect to the system of
quasiparticles (the normal component of liquid or matter, whose flow is
characterised by the normal velocity ${\bf v}_n$),  that there is a force
acting
between the normal component and the vortex. Integration of the anomalous
momentum transfer in Eq.(\ref{MomentoProduction2}) over the cross-section of
the soft core of the moving ATC vortex gives the following force acting on the
vortex (per unit length) from the system of quasiparticles
\cite{Volovik1992}:
\begin{equation}
{\bf F}_{sf}=\int d^2 r{ p_F^3\over {2\pi^2}} \hat {\bf l} ~(
\partial_t \hat {\bf l} \cdot  (\vec\nabla\times \hat {\bf l} ))=-2\pi \hbar
 C_0{\hat {\bf z}}  \times ({\bf v}_L-{\bf v}_n) ,
\label{SpFlowForce}
\end{equation}
where
\begin{equation}
C_0=  p_F^3/3\pi^2~.
\label{C0}
\end{equation}
 Note that this spectral-flow force is
transverse to the relative motion of the vortex and thus is nondissipative
(reversible). In this derivation it was assumed that the quasiparticles and
their momenta, created by the spectral flow from the vacuum, are finally
absorbed by the normal component. The time delay in the process of
absorption and also the viscosity of the normal component lead to a dissipative
(friction) force between the vortex and the normal component:
${\bf F}_{fr}=-\gamma ({\bf v}_L-{\bf v}_n)$. There is no momentum exchange
between the vortex and the normal component if they move with the same
velocity.

Another important
property of the spectral-flow force (\ref{SpFlowForce})
is that it does not depend on the details of
the vortex structure: The result for ${\bf F}_{sf}$ is robust against any
deformation of the
${\hat{\bf l}}$-texture which does not change the asymptote, i.e. the topology
of the vortex. In this respect this force resembles another force, which acts
on the vortex moves with respect to the superfluid  vacuum. This is the
well-known Magnus force:
\begin{equation}
{\bf F}_{M}=  2\pi \hbar
 n_3 {\hat {\bf z}}  \times ({\bf v}_L-{\bf v}_s(\infty)) ~.
\label{MagnusForce}
\end{equation}
Here $n_3$ is the particle density (here the number density of $^3$He
atoms) and
${\bf v}_s(\infty)$ is the uniform velocity of the superfluid vacuum far from
the vortex.

The balance between all the forces acting on the vortex, ${\bf F}_{sf}$, ${\bf
F}_{fr}$, ${\bf F}_{M}$ and some other forces, including any external force
and the so-called Iordanski\v{\i} force coming from the
gravitational analog of the Aharonov-Bohm effect \cite{AB}, determines
the velocity of the vortex  and causes it to  be a linear combination of ${\bf
v}_s(\infty)$ and
${\bf v}_n$.  Due
to this balance the fermionic charge (the linear momentum), which is
transferred
from the fermionic heat bath to the vortex texture, is further transferred from
the vortex texture to the superfluid motion. Thus the vortex texture serves as
intermediate object for the momentum exchange between the fermionic matter and
the superfluid vacuum. In this respect the texture corresponds to the sphaleron
or to the cosmic string in relativistic theories.

The result (\ref{SpFlowForce}) for  the spectral-flow force,
derived  for the ATC vortex from the
axial anomaly equation (\ref{MomentoProduction2}), was confirmed in a
microscopic theory, which took into accout the discreteness of the
quasiparticle spectrum in the soft core
\cite{Kopnin1993}. This was also confirmed in experiments on vortex dynamics in
$^3$He-A
\cite{BevanNature,BevanJLTP}.

In such experiments a uniform array of vortices is produced
by rotating the whole cryostat. In equilibrium the vortices and the normal
component (heat bath) of the fluid  rotate together with the cryostat.
An electrostatically driven vibrating diaphragm  produces an oscillating
superflow, which via the Magnus force generates the vortex motion, while the
normal component remains clamped due to its high viscosity. This creates a
motion
of  vortices with respect both to the heat bath and the superfluid vacuum. The
vortex velocity
${\bf v}_L$  is determined by the overall
balance of forces acting on the vortices, which in the absence of the external
forces can be expressed in terms of the two parameters, so-called mutual
friction parameters:
\cite{BevanNature}
\begin{equation}
\hat{\bf z}\times ({\bf v}_L-{\bf v}_s(\infty))+
d_\perp \hat{\bf z}\times({\bf
v}_n-{\bf v}_L)+d_\parallel({\bf v}_n-{\bf v}_L) =0~~.
\label{ForceBalance}
\end{equation}
Measurement of the damping of the diaphragm resonance and of the
coupling between different eigenmodes of vibrations enables both
parameters, $d_\perp$ and $d_\parallel$, to be deduced.

From the above theory of spectral flow in the $^3$He-A vortex texture it
follows that the parameter, which characterizes the transverse forces acting on
the vortex, is given by
\begin{equation}
d_\perp\approx {C_0-n_3+n_s(T)\over n_s(T)}~,
\label{d_perpAphase}
\end{equation}
where $n_s(T)$
is the density of the superfluid component. In this equation the parameter
$C_0$ from Eq.(\ref{C0}) arises through the axial anomaly, the particle density
$n_3$ stems from the Magnus force and the superfluid density $n_s(T)$ from the
combined effect of Magnus and Iordanski\v{\i} forces. The effect of the chiral
anomaly is crucial for the parameter $d_\perp$   since
$C_0$ is comparable with $n_3$, since $C_0=p_F^3/3\pi^2$ is the particle
density
of liquid $^3$He in the normal state. The difference between $C_0$ and $n_3$
is thus determined by the tiny effect of superfluidity on the particle density
and is extremely small:
$n_3 -C_0\sim n_3 (\Delta_0/v_Fp_F)^2=n_3(c_\perp/c_\parallel)^2 \ll
n_3$. Because of the axial anomaly  one must have
$d_\perp\approx 1$ for all practical temperatures, even including the region
close to $T_c$, where the superfluid component $n_s(T) \sim n_3 (1-T^2/T_c^2)$
is small. $^3$He-A experiments, made  in the whole temperature
range where $^3$He-A is stable,  gave precisely
this value within experimental uncertainty,
$|1-d_\perp|<0.005$\cite{BevanNature}.

This provides an experimental verification of the Adler-Bell-Jackiw axial
anomaly equation (\ref{ChargeParticlProduction}), applied to $^3$He-A, and thus
supports the idea that baryonic charge (and also leptonic charge)
can be generated by electroweak fields.

\subsection{Singular vortex and baryogenesis by cosmic strings}

There are many different scenarios
of the electroweak baryogenesis \cite{Dolgov,Turok}. In some of them the
baryonic charge is created in the cores of topological objects,
in particular in the core of cosmic strings. While a weak and hypercharge
magnetic flux is always present  in the   core of electroweak strings, a weak
and hypercharge electric field can be present along the string if the
string is moving across a background electromagnetic field
\cite{ewitten} or in certain other  processes such as the
de-linking of two linked loops \cite{tvgf,jgtv}. Parallel electric and magnetic
fields in the  string change the baryonic charge and can lead to cosmological
baryogenesis
\cite{barriola} and to the presence of antimatter  in cosmic
rays \cite{gstv}.

Again the axial anomaly is instrumental for the   baryoproduction in the core
of cosmic strings. But now the effect cannot be
described by the anomaly equation (\ref{ChargeParticlProduction}).  This
equation was derived using the energy spectrum of the free massless fermions in
the presence of the homogeneous electric and magnetic fields. But in cosmic
strings these fields are no more homogeneous. Moreover the massless fermions
exist only in the vortex core as bound states in the  potential well produced
by the order parameter (Higgs) field.  Thus the consideration of
baryoproduction should be essentially different: the spectral flow phenomenon
has to  be studied using an exact spectrum of the massless bound
states, namely fermion zero modes on strings.
%%%%%%%%%%%%%%%%%%%%%%%%%%%%%%%%%%%%%%%%%%%%%%%%%%%%%%%%%%%
\begin{figure}[!!!t]
%\centerline{\epsfxsize=0.40\textwidth\epsfbox{Lammi5.eps}}
%\bigskip
\begin{center}
\leavevmode
\epsfig{file=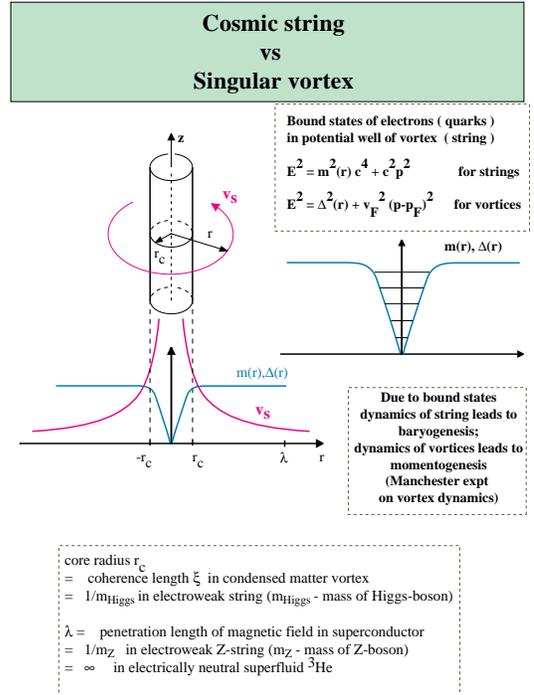,width=0.8\linewidth}
\caption[VortexString]
    {A cosmic string is the counterpart of the Abrikosov vortex in
superconductors. It also supports  bound states of the fermions, which is
important for the electroweak baryoproduction.}
\label{VortexString}
\end{center}
\end{figure}
%%%%%%%%%%%%%%%%%%%%%%%%%%%%%%%%%%%%%%%%%%%%%%%%%%%%%%%%%%%

\begin{figure}[!!!t]
%\centerline{\epsfxsize=0.40\textwidth\epsfbox{Lammi6.eps}}
%\bigskip
\begin{center}
\leavevmode
\epsfig{file=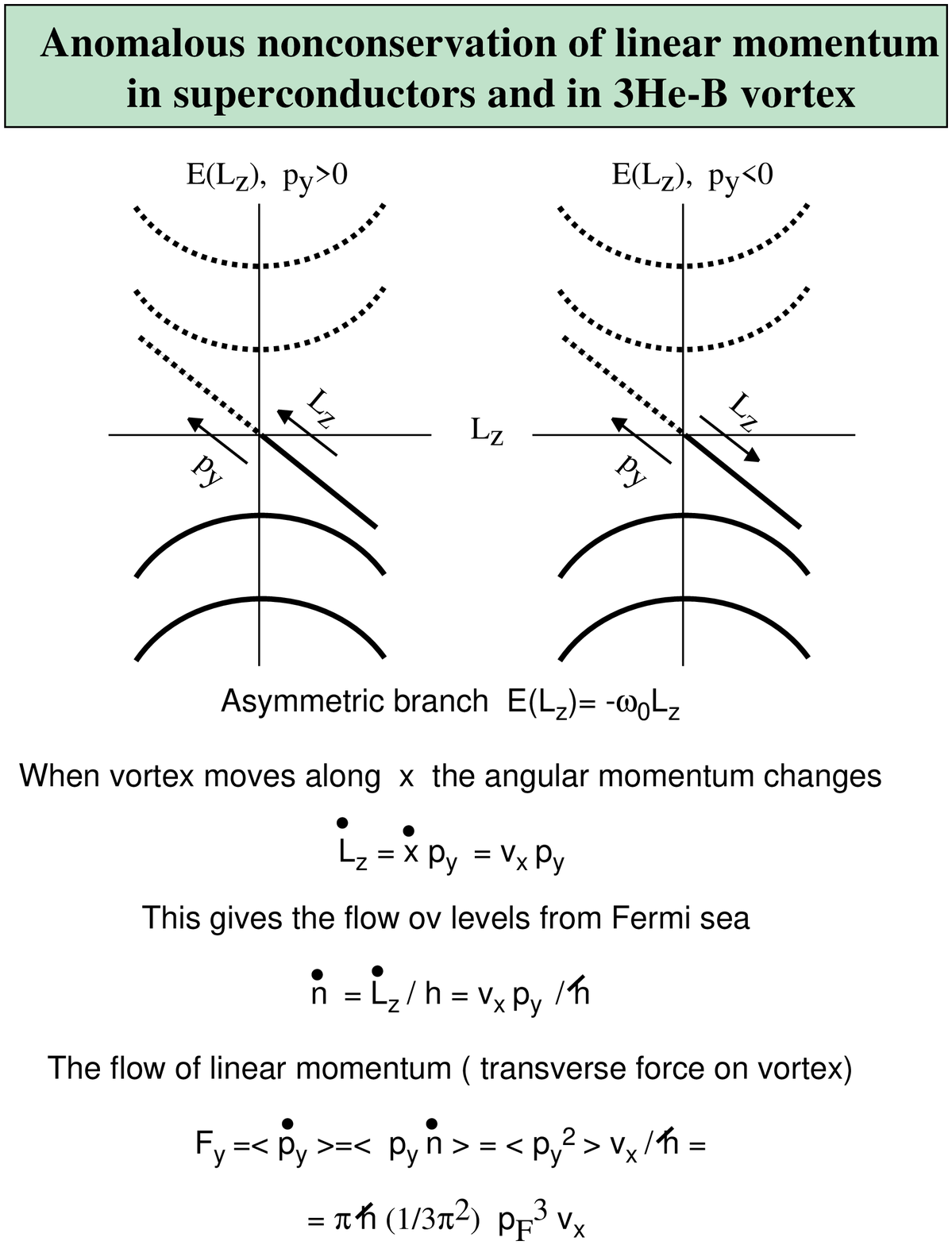,width=0.8\linewidth}
\caption[SingularMomentogenesis]
    {Spectral flow of momentum in the core of
the moving singular vortex leads to the
experimentally observed reactive force on a vortex in superfluid $^3$He-B.}
\label{SingularMomentogenesis}
\end{center}
\end{figure}

A similar situation  takes place in condensed matter, where the counterpart of
the cosmic string is the conventional quantized vortex with a singular core
(Fig.~\ref{VortexString}). The vortices with singular cores are:  (i) Abrikosov
vortices in superconductors; (ii) vortices in superfluid $^3$He-B; and
(iii)  such vortices in
$^3$He-A which, as distinct from the continuous vortices, belong to the
nontrivial elements of the $\pi_1$ homotopy group.  It appears that the
momentogenesis due to the axial anomaly also takes place here, but as distinct
from  the case of the continous ATC vortex in $^3$He-A, it cannot be described
by the continuous anomaly equation of the type of
Eq.~(\ref{MomentoProduction1}). For its description one should consider the
spectral properties of the fermion zero modes localized in the singular vortex
core. The main difference between fermion zero modes in  relativistic strings
and in the conventional condensed matter vortices is the following.  In strings
the anomalous branch
$E(p_z)$  which crosses zero and gives rise to the spectral flow from the
negative vaccum energy  levels  to the positive matter energy levels of
 is given in terms of a continuous variable -- the linear momentum
$p_z$ along the string. In contrast, in the case of condensed-matter
vortices (Fig.~\ref{SingularMomentogenesis}) the
branch $E(L_z)$ is ``crossing'' zero as  a function of the {\em
discrete} angular momentum
$\hbar L_z$ ($L_z$ can be integral or  half-odd integral). The level flow
along the discrete energy levels is suppressed and is determined by the
 interlevel distance
$\hbar\omega_0$ and the level width $\hbar/\tau$ resulting
from the scattering of core excitations by free excitations in
the heat bath outside the core (or by impurities in superconductors).

%%%%%%%%%%%%%%%%%%%%%%%%%%%%%%%%%%%%%%%%%%%%%%%%%%%%%%%%%%%
%%%%%%%%%%%%%%%%%%%%%%%%%%%%%%%%%%%%%%%%%%%%%%%%%%%%%%%%%%

This suppression of spectral flow  results in a renormalization of
the spectral-flow parameter, which is roughly
\cite{KopninVolovikParts,StoneSpectralFlow}
\begin{equation}
 \tilde C_0 \sim {C_0\over 1+\omega_0^2\tau^2}~,
\label{C_=Renormalized}
\end{equation}
and the Eq. (\ref{d_perpAphase}) becomes:
\begin{equation}
d_\perp\approx {\tilde C_0-n_3+n_s(T)\over n_s(T)}
\label{d_perpBphase}
\end{equation}
If
$\omega_0\tau \ll  1$, the levels overlap and spectral flow is allowed. In
the opposite limit $\omega_0\tau \gg  1$ it is completely suppressed. The
parameter $\omega_0\tau $ depends on temperature and this allows us to check
Eq.~(\ref{d_perpBphase}) experimentally. This has been done in an
experiment in Manchester on the dynamics of singular vortices in
$^3$He-B  \cite{BevanNature,BevanJLTP}. An equation of the type of
(\ref{d_perpBphase}) has been verified in a broad temperature range, which
included both extreme limits, $\omega_0\tau \ll  1$ and $\omega_0\tau \gg  1$.

\section{MAGNETIC FIELD from FERMIONIC CHARGE}

A recent scenario of the generation of primordial magnetic fields
by Joyce and
Shaposhnikov \cite{JoyceShaposhnikov,GiovanniniShaposhnikov}  is based on an
effect, which is the inverse to that discussed in the previous section.  The
axial anomaly gives rise to a transformation of
an excess of chiral particles into a
hypermagnetic field.   In $^3$He-A language this process
describes the collapse of excitation momenta (fermionic charges) towards the
formation of textures. These textures are the
counterpart of the hypermagnetic field in the
Joyce-Shaposhnikov scenario
\cite{NaturePrimordial}  (Figs.~\ref{Counterflow},\ref{PrimordialField}).  Such
a collapse of quasiparticle momentum
was recently observed in  the rotating cryostat of the Helsinki Low
Temperature Laboratory
\cite{Experiment,NaturePrimordial}.

\subsection{Effective Lagrangian at low $T$}

To study the instability of the superflow and relate it to the problem of
magnetogenesis, let us start with the relevant Effective Lagrangian for
superfluid dynamics at low $T$ and find the correspondence to the
effective Lagrangian for the system of chiral fermions interacting with the
magnetic or hypermagnetic field via the axial anomaly conversion process. For
the hydrodynamic action in  $^3$He-A we shall use the known results
collected in
the book \cite{VollhardtWolfle}.

Consider
a superfluid moving with respect to the  walls of container. The normal
component
of the liquid  is clamped by the vessel walls due to its high
viscosity, so that
the normal velocity ${\bf v}_n=0$ in the reference frame moving with the
vessel. If the superfluid velocity
${\bf v}_s$ of the condensate in Eq.~(\ref{v_s}) is nonzero in this reference
frame, one has a nonzero counterflow of the superfluid and normal components
with relative velocity ${\bf w}={\bf v}_s - {\bf v}_n$. This relative
velocity provides a nonzero fermionic charge of matter, as will be seen below,
and the flow instability leads to the transformation of this charge to the
analogue of the hypermagnetic field.   Let us choose
the axis $z$ along the  velocity ${\bf w}$ of the
counterflow. In
equilibrium the unit orbital vector ${\hat{\bf l}}$ is oriented along the
counterflow: ${\hat{\bf l}}_0= {\hat{\bf z}}$. The stability problem is
investigated using the quadratic form of the deviations of the
superfluid velocity and the
${\hat{\bf l}}$-vector from their equilibrium values:
\begin{equation}
{\bf {\hat l}} = {\bf {\hat l}}_0 + \delta {\bf {\hat l}}({\bf r},t)
-{1\over 2}{\bf {\hat l}}_0 (\delta {\bf {\hat l}}({\bf r},t))^2~,~ {\bf v}_s =
{\bf w}_0 + \delta {\bf v}_s({\bf r},t)\ .
\label{oscillating-l}
\end{equation}
The instability of the counterflow towards generation
of the inhomogenity $\delta {\bf {\hat l}}({\bf r},t)$,   corresponds to
the generation
of the magnetic field ${\bf B}=p_F{\bf \nabla}\times \delta {\hat {\bf
l}}$ from the chiral fermions.

There are 3 terms in the energy of the liquid, which are relevant for our
consideration of stability of superflow at low
$T$:
\begin{eqnarray}
\nonumber
F={1\over 2}m_3n_s^{ij}v_{si}v_{sj} + C_0 ({\bf v}_s\cdot {\bf {\hat
l}})({\bf {\hat l}}\cdot (\nabla\times{\bf {\hat l}}))\\
 +K_b({\bf {\hat
l}}\times (\nabla\times{\bf {\hat l}}))^2
\label{F}
\end{eqnarray}
(1) The first term in Eq.~(\ref{F}) is the kinetic energy of superflow with
$n_s^{ij}$ being the anisotropic tensor of superfluid density. At low $T$
one has
\begin{equation}
n_s^{ij}\approx  n_3 \delta^{ij}- n_{n\parallel}{\hat
l}^{i}{\hat l}^{j}~,~ n_{n\parallel}
\approx { m^*\over 3m_3} p_F^3
{T^2\over \Delta_0^2} ~~.
\label{LowTnormalDensity}
\end{equation}

(2) The second term in Eq.~(\ref{F}) is the anomalous interaction of the
superflow with the ${\hat{\bf l}}$-texture, coming
from the axial anomaly
\cite{exotic}. The anomaly parameter $C_0$ at
$T=0$ is the same as in  Eq.~(\ref{C0}).

(3) Finally the third term
is the relevant part of the
energy of ${\hat{\bf l}}$-texture. There are two other terms in
the textural energy \cite{VollhardtWolfle}, containing  $({\bf {\hat l}}\cdot
(\nabla\times{\bf {\hat l}}))^2$ and $(\nabla\cdot{\bf {\hat l}})^2$, but they
are not important for the stability problem: The instability starts when the
$z$-dependent disturbances begin to grow. Therefore we are interested only in
$z$-dependent
${\hat{\bf l}}$-textures, which in a quadratic approximation contribute
the term $({\bf {\hat l}}\times
(\nabla\times{\bf {\hat l}}))^2$. The rigidity $K_b$ at low $T$ is
logarithmically divergent
\begin{equation}
K_b=
{{p_F^2v_F}\over {24\pi^2\hbar}}~{\rm ln}~ \left ( {\Delta_0^2\over T^2 }~
\right )~~,
\label{Kb}
\end{equation}
which we shall later relate to the zero charge effect in relativistic theories
\cite{exotic}.

There is also a topological connection between ${\bf {\hat l}}$ and  ${\bf
v}_s $, since ${\bf v}_s$ in Eq.~(\ref{v_s}) represents torsion of the dreibein
 ${\hat{\bf e}}_1,{\hat{\bf e}}_2,{\hat{\bf e}}_3$ field. This leads to a
nonlinear connection, the so-called Mermin-Ho relation \cite{VollhardtWolfle},
which in our geometry gives
\begin{equation}
\delta {\bf v}_s = {\hbar\over 2m_3} {\bf {\hat z}}\partial_z \Phi +
{\hbar\over 4m_3}   \delta {\bf {\hat l}}\times
\partial_z
\delta{\bf {\hat l}}
\label{MerminHo}
\end{equation}
 The three variables, the  potential $\Phi$ of the flow velocity and
the two components $\delta {\bf {\hat l}}\perp  {\bf {\hat l}}_0$ of the unit
vector ${\bf {\hat l}}$, are just another presentation of 3 rotational
degrees of
freedom of the dreibein ${\hat{\bf e}}_1,{\hat{\bf e}}_2,{\hat{\bf e}}_3$
 (the
rotation of vectors
${\bf m}=c_\perp {\hat{\bf e}}_1$, ${\bf n}=c_\perp {\hat{\bf e}}_2$ and
$\hat l$ in Fig.~\ref{CollectiveModes}).
Whereas  $\delta {\bf {\hat l}}$ is responsible for the effective
vector potential of the (hyper) magnetic field, the variable $\Phi$
-- the angle of
rotation of vectors
${\bf m}=c_\perp {\hat{\bf e}}_1$ and ${\bf n}=c_\perp {\hat{\bf e}}_2$ about
axis $\hat l$ in Fig.~\ref{CollectiveModes} -- represents an {\em axion} field
as we shall see later.

Let us expand the energy in terms of small
perturbations $\delta\hat{\bf l}$.
Adding terms with time derivatives we obtain the following
Lagrangian for $\Phi$  and $\delta {\bf {\hat l}}$:
\begin{equation}
L=F_0+ L_{\delta {\bf {\hat l}}}+ L_{\Phi} ~.
\label{L}
\end{equation}
Here $F_0$ is the initial homogeneous flow energy
\begin{equation}
F_0={1\over 2}m_3n_3{\bf w}_0^2 - { m^*\over 6m_3} p_F^3
{T^2\over \Delta_0^2}({\bf w}_0 \cdot {\bf {\hat l}}_0)^2 \,,
\label{F0}
\end{equation}
and
$ L_{\delta {\bf {\hat l}}}$ is the textural Lagrangian
of order $(\delta\hat{\bf l})^2$:
\begin{eqnarray}
 L_{\delta {\bf {\hat l}}}={{p_F^2}\over {24\pi^2\hbar v_F}}~{\rm ln} \left (
{\Delta_0^2\over T^2 }~
\right )~ \left[v_F^2(\partial_z \delta{\bf {\hat l}})^2 - (\partial_t
\delta{\bf {\hat l}})^2\right]
\label{Ll1}\\
 + ~{p_F^3 \over 2\pi^2}~({\hat{\bf
l}}_0\cdot{\bf w}_0) (\delta{\hat{\bf l}}\cdot {\bf\nabla}\times
\delta{\hat{\bf
l}})
\label{Ll2}\\
+~{ m^*\over 6} p_F^3
{T^2\over \Delta_0^2}({\bf w}_0 \cdot {\bf {\hat l}}_0)^2~(\delta{\hat{\bf
l}})^2
\label{Ll3}
\end{eqnarray}
The first term, Eq.(\ref{Ll1}), describes the propagation of  textural waves
(the so-called orbital waves which play the part of the hyperphoton, see
below). The Eq.(\ref{Ll3}) gives the mass of the hyperphoton. The term in
Eq.(\ref{Ll2}) is the Chern-Simons term in action (see below) which is the
consequence of the axial anomaly and thus contains the same  factor
${p_F^3
\over 2\pi^2}=(3/2)C_0$ as in Eq.~(\ref{MomentoProduction2}). To obtain
this factor from the hydrodynamic action for $^3$He-A one should collect all
the relevant terms: (i) The factor
$C_0$ comes from Eq.~(\ref{F}). (ii) The factor $n_3/2$ --
from Eq.(\ref{MerminHo}). And (iii) the factor $-(n_3-C_0)/2$ -- from the
intrinsic angular momentum. Altogether they give
$C_0 +n_3/2 - (n_3-C_0)/2=(3/2)C_0=p_F^3
/ 2\pi^2$ in Eq.~(\ref{Ll2}). We do not discuss  the problem of intrinsic
angular momentum, though it is clearly related to the axial anomaly and
spectral flow \cite{OrbitalMomentum}. Here it is important that
the contribution  of the intrinsic
angular momentum to the
hydrodynamic action is \cite{exotic}
\begin{equation}
{1 \over 2}~(n_3-C_0)~\left({\hat{\bf
l}}_0\cdot (\delta{\hat{\bf l}}\times (\partial_t +
 {\bf w}\cdot {\bf\nabla})
\delta{\hat{\bf l}})\right)
 ~,
\label{IntrinsicMomentum}
\end{equation}
and this gives the required factor $-(n_3-C_0)/2$.

$  L_{\Phi}$ is the variation of the Lagrangian for superflow. At low $T$ one
has
\begin{eqnarray}
  L_{\Phi}={\hbar^2\over 8 m_3}n_3  \left[(\partial_z \Phi)^2 -
{1\over s^2}(\partial_t
\Phi)^2 \right]
\label{Lphi1}\\
 + ~{3\hbar\over 4m_3} C_0 ~ \partial_z \Phi~ (\delta{\hat{\bf l}}\cdot
{\bf\nabla}\times
\delta{\hat{\bf l}})
\label{Lphi2}
\end{eqnarray}
The first two terms of this Lagrangian, contained in
Eq.~(\ref{Lphi1}), describe the propagation of sound waves (phonons), and $s$
is the
speed of sound. We shall later relate the sound waves to axions,
because of their coupling with the density of topological charge in
Eq.~(\ref{Lphi2}).

Let us now establish all these correspondences step by step.

\subsection{Fermionic  charge and Chern-Simons energy}

\begin{figure}[!!!t]
%\centerline{\epsfxsize=0.40\textwidth\epsfbox{Lammi7.eps}}
%\bigskip
\begin{center}
\leavevmode
\epsfig{file=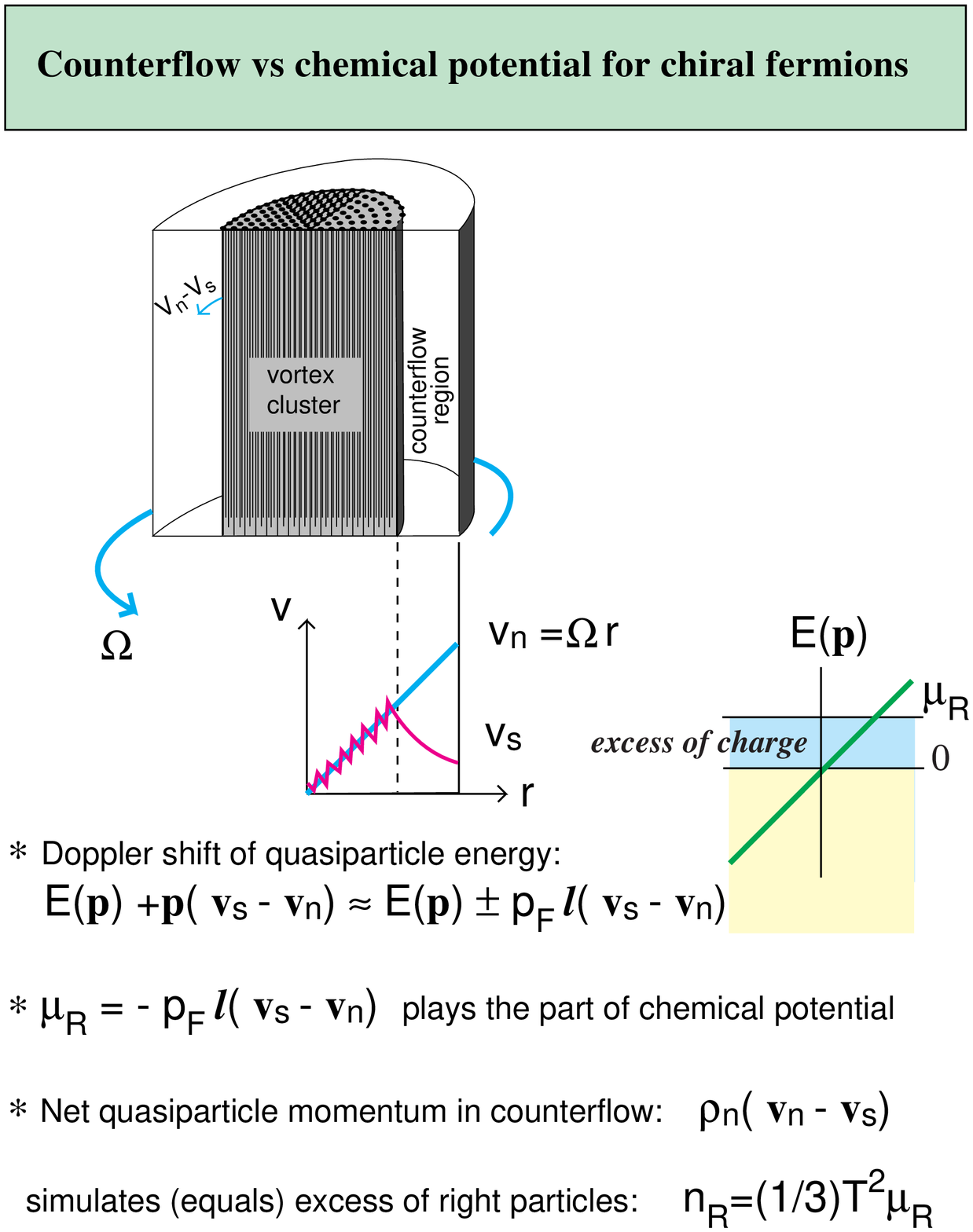,width=0.8\linewidth}
\caption[Counterflow]
    { The counterflow generated by the rotation of the cryostat is
equivalent to the nonzero chemical potential for the right electrons.
If the
cryostat rotates, the typical situation is that the vortices form
a vortex cluster in the center of the vessel. Within the cluster the
average superfluid velocity $<{\bf
v}_s>$ follows the solid body-like rotation velocity of the normal
component: $<{\bf
v}_s>= {\bf v}_n  = {\bf\Omega}\times {\bf r} $. Outside the cluster the
normal component performs  a solid body rotation, while the superfluid is
irrotational. There results a  counterflow
${\bf w}={\bf v}_s - {\bf
v}_n$ between the superfluid and normal components. In the counterflow region
the
${\hat{\bf l}}_0$-vector is oriented along the counterflow. The Doppler
shift of the quasiparticle energy plays the part of   the
chemical potential
$\mu_R$ and $\mu_L$ for the chiral fermions in the vicinity of two gap nodes.
In the counterflow  the quasiparticles have net linear momentum. This excess of
the fermionic charge of quasiparticles is equivalent to the excess of chiral
right-handed electrons if their chemical potential $\mu_R$ is nonzero.}
\label{Counterflow}
\end{center}
\end{figure}

In the presence of counterflow, ${\bf w} ={\bf v}_s-{\bf
v}_n$, of the motion of
the superfluid component of $^3$He-A  with respect to the
normal fraction,  the energy of quasiparticles is Doppler shifted by an amount
${\bf p}\cdot{\bf w}$, which is $\approx \pm p_F({\hat{\bf l}}_0\cdot{\bf
w}_0)$ near the nodes. The counterflow therefore produces an
effective chemical potential for the relativistic fermions in the vicinity of
both nodes (Fig.~\ref{Counterflow}):
\begin{equation}
\mu_R= - p_F({\hat{\bf l}}_0\cdot{\bf w}_0) ~,~\mu_L=-\mu_R~.
\label{ChemicalPotentials}
\end{equation}

%%%%%%%%%%%%%%%%%%%%%%%%%%%%%%%%%%%%%%%%%%%%%%%%%%%%%%%%%%%
%%%%%%%%%%%%%%%%%%%%%%%%%%%%%%%%%%%%%%%%%%%%%%%%%%%%%%%%%%

According to our analogy the relevant fermionic charge of our system, which is
anomalously conserved and which corresponds to the number of right fermions,
is the  momentum of quasiparticles along
${\hat{\bf l}}$ divided by $p_F$. Since the momentum density of
quasiparticles is ${\bf P}= - n_{n\parallel} {\bf w}$, the density of the
fermionic charge is
\begin{equation}
 {P\over p_F}= - {n_{n\parallel}\over p_F} {\hat{\bf l}}_0\cdot{\bf w}_0 ~.
\label{FermionCharge1}
\end{equation}
Using Eq.~(\ref{LowTnormalDensity}) for $n_{n\parallel}$,
Eq.~(\ref{SpeedsInAphase}) and Eq.(\ref{3speeds}) for the metric tensor, and
Eq.~(\ref{ChemicalPotentials}) for the chemical potential, one obtains a very
simple covariant expression for the density of the fermionic charge
\begin{equation}
n_R\equiv{P\over
p_F}={1\over 3} T^2 \mu_R \sqrt{-g}~.
\label{FermionCharge2}
\end{equation}
Here $g$ is the determinant of the metric tensor $g_{\mu\nu}$
\begin{equation}
\sqrt{-g}={1\over c_\parallel
c_\perp^2}={m^*p_F\over \Delta_0^2}~~.
\label{DetG}
\end{equation}

Eq.~(\ref{FermionCharge2}) represents the number density of
chiral right-handed massless electrons induced by the chemical potential
$\mu_R$
at temperature $T$. This is the starting point of the Joyce-Shaposhnikov
scenario of magnetogenesis.  It is assumed there that at an early stage of the
universe, possibly at the Grand Unification epoch ($10^{-35}$ s after the big
bang), an excess of  chiral right-handed electrons,
$e_R$, is somehow produced due to parity violation.

The equilibrium relativistic
energy of the system of right electrons also appears to be completely
equivalent to the kinetic energy of the quasiparticles in the counterflow in
Eq.~(\ref{F0})
\begin{equation}
\epsilon_R =  {1\over 6}
T^2 \mu_R^2\sqrt{-g} \equiv {1\over 2} m_3 n_{n\parallel}({\bf w}_0 \cdot {\bf
{\hat l}}_0)^2 ~,
\label{FermionEnergy}
\end{equation}
The difference in the sign between Eqs.~(\ref{F0}) and  (\ref{FermionEnergy})
is the usual difference between the thermodynamic potentials at fixed
chemical potential and at fixed particle number (fixed velocity and fixed
momentum correspondingly).

Due to the ``inverse'' axial anomaly
the leptonic charge (excess of right electrons)
can be transferred to the ``inhomogeneity'' of the vacuum. This inhomogeneity,
which absorbs the fermionic charge, arises as a hypermagnetic field
configuration. Thus the charge absorbed by the hypermagnetic field,
${\bf\nabla}\times {\bf A}$, can be expressed in terms of its helicity,
\begin{equation}
n_R \{{\bf A}\}={1\over 2\pi^2} {\bf A}\cdot ({\bf\nabla}\times {\bf A})~~.
\label{anomaly}
\end{equation}
The right-hand side is the so called Chern-Simons (or topological) charge
of the magnetic field.

\begin{figure}[!!!t]
%\centerline{\epsfxsize=0.40\textwidth\epsfbox{Lammi8.eps}}
%\bigskip
\begin{center}
\leavevmode
\epsfig{file=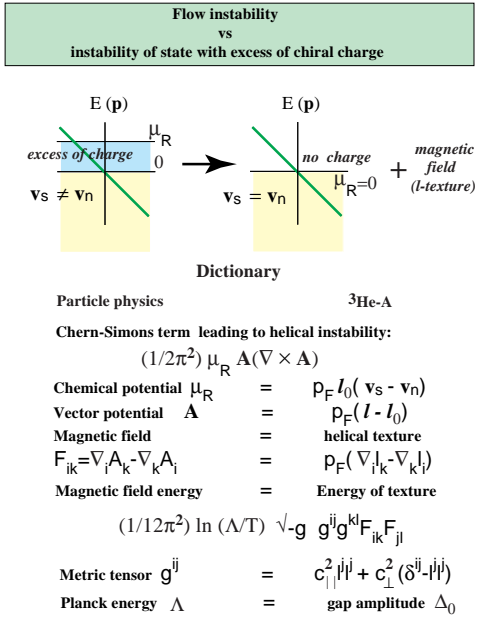,width=0.8\linewidth}
\caption[PrimordialField]
    {An excess of
chiral right-handed electrons in the early Universe can be effectively
converted to a hypermagnetic field via the mechanism of chiral anomaly.
This is described essentially by the same equations as the counterflow
instability observed in $^3$He-A.}
\label{PrimordialField}
\end{center}
\end{figure}

When this charge is transformed from the fermions to the hypermagnetic field,
the energy stored in the fermionic system decreases. This leads to a
energy gain which is  equal to the
Chern-Simons charge multiplied by the chemical potential:
\begin{equation}
F_{CS}= n_R \{{\bf A}\} \mu_R=
{1\over 2\pi^2} \mu_R  {\bf A}\cdot ({\bf\nabla}\times {\bf A})  ~~.
\label{csenergy1}
\end{equation}
The translation to the language of
$^3$He-A, according to the dictionary in
Fig.~(\ref{PrimordialField}), gives the following energy change, if the texture
is formed from the counterflow,
\begin{equation}
F_{CS}= {p_F^3\over 2\pi^2}({\hat{\bf
l}}_0\cdot{\bf w}_0) (\delta{\hat{\bf l}}\cdot {\bf\nabla}\times
\delta{\hat{\bf
l}}) ~~.
\label{csenergy2}
\end{equation}
This exactly coincides with Eq.~(\ref{Ll2}).

%%%%%%%%%%%%%%%%%%%%%%%%%%%%%%%%%%%%%%%%%%%%%%%%%%%%%%%%%%%
%%%%%%%%%%%%%%%%%%%%%%%%%%%%%%%%%%%%%%%%%%%%%%%%%%%%%%%%%%

The Chern-Simons term in
Eqs.~(\ref{csenergy1},\ref{csenergy2}) can have arbitrary sign. It is positive
if the conterflow is increased and negative if the counterflow is reduced.
Thus one can have an energy gain from the transformation of the counterflow
(fermionic charge) to the texture (hypermagnetic field). This energy gain is
however to be compared with the positive energy terms in Eq.~(\ref{Ll1}) and
Eq.~(\ref{Ll3}). Let us consider these two terms in more detail.

\subsection{Maxwell Lagrangian for hypermagnetic and hyperelectric fields.}

The   Lagrangian
for the
$\delta{\hat{\bf l}}$-texture in Eq.~(\ref{Ll1}) is completely
equivalent  to the conventional Maxwell Lagrangian for the (hyper-) magnetic
and electric fields. For example the textural energy, written
in covariant form, corresponds to the magnetic energy:
\begin{eqnarray}
F_{\rm magn}= {\rm ln}~ \left ( {\Delta_0^2\over T^2 }~ \right )
{{p_F^2v_F}\over {24\pi^2\hbar}}~
(\partial_z \delta \hat {\bf l})^2
\label{magenergy1}\\
~~\equiv
{{\sqrt{-g}}\over {2\gamma^2} }g^{ij}g^{kl}F_{ik}F_{jl}~~.
\label{magenergy2}
\end{eqnarray}
Here, $F_{ik}= \nabla_i A_k -\nabla_k A_i $, and $\gamma^2$ is a running
coupling constant, which is logarithmically divergent because of vacuum
polarization in a complete analogy with the fine structure constant $e^2/4\pi
\hbar c$:
\begin{equation}
\gamma^{-2} ={1\over 12\pi^2} {\rm ln}~ \left ({\Delta_0^2\over T^2 }~ \right
) ~.
\label{RunningCoupling}
\end{equation}
Eq.~(\ref{magenergy2})
transforms to  Eq.~(\ref{magenergy1}) if one takes into account that in our
geometry the ``hypermagnetic'' field ${\bf B}\perp {\hat{\bf l}}_0$.

The
gap amplitude $\Delta_0$, constituting
the ultraviolet cut-off in the logarithmically
divergent magnetic energy, plays the part of the Planck energy scale. Note that
$\Delta_0$ has a parallel with the Planck energy in some other
situations, too.
For example the analogue of the
cosmological constant, which arises in the effective
gravity of $^3$He-A, has the value
$\Delta_0^4/12\pi^2$ \cite{Volovik1986}.

\subsection{Mass of hyperphoton}

The ``hyperphoton'' in $^3$He-A has a mass. There are several sources of this
mass.

(i) The  value of the
mass of the ``hyperphoton'' is seen from Eq.~(\ref{Ll3}), if  it is
 written in covariant form:
\begin{equation}
F_{mass} (T,\mu_R)={1\over 6} \sqrt{-g} g^{ik} A_i A_k {T^2 \mu_R^2\over
\Delta_0^2}  ~~.
\label{MassEnergy1}
\end{equation}
Thus the mass is
\begin{equation}
M_{ph}^2={\gamma^2\over 3}~ {T^2 \mu_R^2 \over \Delta_0^2}
~~,
\label{Mass1}
\end{equation}
 In $^3$He-A this mass is physical, though
it contains the ``Planck'' energy cut-off $\Delta_0$:
The ``hyperphoton mass'' is
the gap in the spectrum  of orbital waves, propagating oscillations of
$\delta \hat{\bf l}$, which correspond just
to the hyperphoton. This mass appears
due to the presence of counterflow, which provides
the restoring force for  oscillations of $\delta \hat{\bf l}$.

For the relativistic counterpart of
$^3$He-A, the Eq.~(\ref{Mass1}) suggests that the mass of the
hyperphoton could arise if both the temperature
$T$ and the chemical potential $\mu_R$ are finite. Of course, in the case of
exact local $U(1)$ symmetry, the mass of the hyperphoton should be zero. But in
an effective theory, the local $U(1)$ symmetry appears only in the low-energy
corner and thus is approximate. It can be violated (not spontaneously) at
higher energy leading to a nonzero  hyperphoton mass which depends on the
cut-off parameter. And in fact the mass in Eq.~(\ref{Mass1}) disappears in the
limit of an infinite cut-off parameter or is small, if the cut-off is of Planck
scale.  The $^3$He-A thus provides an  illustration of  how the
terms of order $(T/E_{\rm Planck})^2$ appear in the effective quantum field
theory
\cite{Jegerlehner}.

(ii) In the collisionless regime $\omega\tau \gg 1$, a nonzero mass term
is present even in the  absence of the counterflow, ${\bf w}=0$.  It
corresponds
to the high-frequency photon mass in the relativistic plasma, calculated by
Weldon
\cite{Weldon}:
\begin{equation}
 M^2_{ph}(\omega\tau \gg 1)={N_F\over 18}~ \gamma^2 T^2.
\label{Mass2rel}
\end{equation}   Here $N_F$ is the number of
fermionic species and $\gamma$ again is the running coupling constant. This can
be easily translated to  $^3$He-A language, since mass
is a covariant quantity.  Substituting the running coupling from
Eq.~(\ref{RunningCoupling}) and taking into account that the number of the
fermionic species in $^3$He-A is $N_F=N_{FR} +N_{FL}= 2$, one obtains the
gap in
the spectrum of the high-frequency orbital waves (called also the normal
flapping
mode)
\begin{equation}
 M^2_{\rm orb~waves}(\omega\tau \gg 1)={4\pi^2\over 3}  {T^2 \over \ln
(\Delta_0^2/T^2)}~.
\label{Mass2Aphase}
\end{equation}
This coincides with  Eq.~(11.76b) of
Ref.\cite{VollhardtWolfle} for the normal flapping mode. Note that in $^3$He-A
this gap in the spectrum, corresponding to the relativistic plasma
oscillations, was obtained by W\"olfle already in 1975 \cite{Wolfle}.

The corresponding mass term in the Lagrangian for the gauge bosons is
\begin{equation}
 F_{mass} (T,\omega\tau \gg 1)= {N_F\over 36} T^2  \sqrt{-g}
g^{ik}A_iA_k~~.
\label{MassEnergy2}
\end{equation}
which is valid both for the proper
relativistic theory with chiral  fermions and
for $^3$He-A, where
$N_F=2$.

 (iii) There is also the topological mass of the ``photon'' in $^3$He-A, which
comes from the axial anomaly and intrinsic angular momentum
\cite{exotic,Volovik1975,LeggettTakagi}. It is rather small.

(iv) The tiny mass
coming from the spin-orbital interaction in
$^3$He-A \cite{LeggettTakagi} is described by the energy term
\cite{VollhardtWolfle}
\begin{equation}
-g_D(\hat {\bf l}\cdot \hat {\bf d})^2~,
\label{SpinOrbit}
\end{equation}
 where $\hat {\bf d}$ is the unit vector of the
spontaneous  anisotropy in spin space. This term has no counterpart in
relativistic theories but is important in NMR experiments on $^3$He-A (see
below).

Here we discussed how the ``photon'' mass in $^3$He-A is influenced by
variuos external and internal factors: counterflow (chemical potential),
temperature, anomaly, spin-orbital interaction, Planck cut-off parameter. These
factors also influence the speed of ``light'' in $^3$He-A and this occurs
essentially in the same manner as in relativistic theories (see
\cite{DittrichGies} for references on the
modification of the speed of light by  electromagnetic fields, temperature,
gravitational background, and other external environments).  The only
difference is that in $^3$He-A the environment modifies the Planck cut-off
parameter as well \cite{GravitationalConstant}, which gives an extra dependence
of the ``photon'' mass  and the speed of ``light'' on the environment.

\subsection{Instability towards magnetogenesis}

For us the most important property of the axial anomaly term in
Eqs.~(\ref{csenergy1},\ref{csenergy2}) is that it is linear in the
derivatives of
$\delta {\bf {\hat l}}$. Its sign thus can be negative, while its magnitude can
exceed the positive quadratic term in eq. (\ref{magenergy1}). This
leads  to the
helical instability towards formation of the inhomogeneous $\delta {\bf {\hat
l}}$-field. During this instability the kinetic energy of the
quasiparticles in the counterflow (analogue of the energy stored in the
fermionic
degrees of freedom) is converted into the energy of the inhomogeneity
$ {\nabla}_z
\delta{\hat{\bf l}}$, which is the analogue of the magnetic energy of the
hypercharge field.

This instability can be found by investigation of the eigenvalues of
the quadratic form describing the energy in terms of  ${\bf
A}=p_F\delta{\hat{\bf l}}$ in Eq.(\ref{Ll1}-\ref{Ll3}). Using the covariant
form of this equation one obtains the following $2\times
2$ matrix for  two components of the vector potential,
$A_x=A_{x0}e^{iqz}$ and $A_y=A_{y0}e^{iqz}$ :
\begin{equation}
\left( \matrix
{M_{ph}^2 + c_\parallel^2q^2
& {\gamma^{2}\over 2\pi^2}\mu_R c_\parallel q \cr {\gamma^{2}\over 2\pi^2}\mu_R
c_\parallel q & M_{ph}^2  + c_\parallel^2q^2 \cr }
\right) ~~.
\label{Matrix}
\end{equation}

This matrix is applied both to the
Joyce-Shaposhnikov scenario and to the instability
of the $^3$He-A superflow.
This is one of the rare cases when the equation of motion
for the $\hat {\bf l}$-vector reduces to relativistic (Maxwell + Chern-Simons)
equations.  This stems from the fact that for the investigation of the
stability one needs an energy which is quadratic in terms of the small
deviations of the vector potential (vector $\hat {\bf l}$) from the uniform
background. In our geometry:  (i) The equilibrium unit vector $\hat {\bf l}_0$
is oriented in one direction (along the velocity), which means that the
background metric is constant in space. (ii) Small deviations $\delta \hat {\bf
l} \equiv {\bf A}/p_F$  of the vector $\hat {\bf l}$ from equilibrium are
perpendicular to the flow, while the relevant coordinate dependence
(i.e. that which
leads to instability) is the $z$-dependence along the flow. Thus there are no
derivatives in $x$ and $y$ in the relevant Lagrangian, while ${\bf A}$ contains
only the transverse components. (iii) The Lagrangian is quadratic in
the gauge
field ${\bf A}\equiv  p_F \delta  \hat {\bf l}$, while the metric enters
only as
a constant (though anisotropic) background. All these facts
conspire to produce a
complete analogy with the relativistic theory.  Such a geometry, in which the
analogy is exact, is really unique, and it might be called a
 miracle that it indeed does occur in a
real experimental situation.

The quadratic form in Eq.~(\ref{Matrix}) becomes negative if
\begin{equation}
 {\mu_R\over M_{ph}}> {4\pi^2\over \gamma^2  } ~~.
\label{StabilityCondition1}
\end{equation}
Inserting the photon mass from Eq.~(\ref{Mass1}), one finds that the uniform
counterflow becomes unstable towards the nucleation of the texture if
\begin{equation}
 {T\over \Delta_0} \ln^{1/2} \left({\Delta_0^2\over T^2}\right)< {3\over
2\pi  } ~~.
\label{StabilityCondition2}
\end{equation}
If this condition is fulfilled, the instability occurs for any value of the
counterflow (any value of the chemical potential $\mu_R$ of right electrons).

In relativistic theories, where $\Delta_0$ is the Planck energy, this
condition is always fulfilled. Thus the excess of the fermionic charge is
always
unstable towards nucleation of the hypermagnetic field. In the scenario of the
magnetogenesis developed by Joyce and Shaposhnikov
\cite{JoyceShaposhnikov,GiovanniniShaposhnikov}, this instability is
responsible for the genesis of the hypermagnetic field well above the
electroweak transition. The role of the subsequent electroweak transition is to
transform this hypermagnetic field  to the conventional
(electromagnetic $U(1)$) magnetic field due to the electroweak symmetry
breaking.

In
$^3$He-A the Eq.(\ref{StabilityCondition2}) shows that the instability
always occurs if the temperature is low enough compared to $\Delta_0$ (
$\Delta_0 \sim 2 T_c$. What happens at $T \sim T_c$ is not clear from
Eq.(\ref{StabilityCondition2}), since our analysis works only in
the limit
$T\ll
\Delta_0$. So, the rigorous  theory is required, which
holds at any $T$. The helical
instability in $^3$He-A  has been intensively discussed theoretically
(see, e.g., \cite{ThesisVollhardt}). According to a rigorous  theory, which
takes into account the Fermi-liquid
parameters, the counterflow is unstable at any $T$ if the spin-orbital
coupling in Eq.~(\ref{SpinOrbit}) is neglected, i.e. $g_D=0$, but is stable
at $T$ above about
$0.8 T_c$ if the spin-orbital
coupling is taken into account and the stiffness of the spin vector $\hat
{\bf d}$ suppresses the instability (see Sec. 7.10.1 in the book
\cite{VollhardtWolfle}).  The result of the helical instabilty can be
either the
formation of the helical texture with small opening angle or the complete
collapse of the counterflow.  In the first case only some part of the
counterflow momentum transforms to the momentum of the helix. In the second
case  the collapse of the counterflow leads to the formation of continuous ATC
vortices and thus the whole counterflow momentum is transformed to the momentum
carried by the vortex texture. Experimentally the second scenario, with
formation of vortices, is realized \cite{Experiment}.

%%%%%%%%%%%%%%%%%%%%%%%%%%%%%%%%%%%%%%%%%%%%%%%%%%%%%%%%%%
\begin{figure}[!!!t]
%\centerline{\epsfxsize=0.40\textwidth\epsfbox{Lammi9.eps}}
%\bigskip
\begin{center}
\leavevmode
\epsfig{file=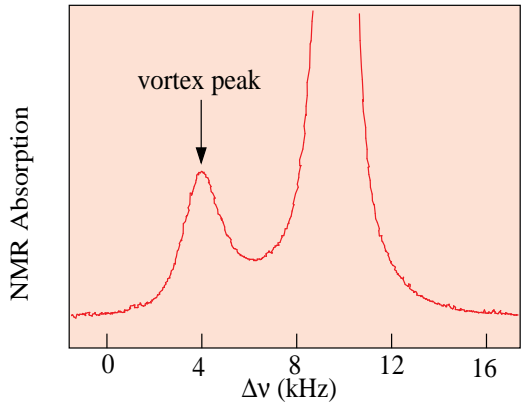,width=0.8\linewidth}
\caption[experiment]
    {The NMR signal from  array of ATC vortices in the container.
The position of the satellite peak indicates the type of the vortex, while the
intensity is proportional to the number of vortices  of this type in the cell.}
\label{experiment}
\end{center}
\end{figure}
%%%%%%%%%%%%%%%%%%%%%%%%%%%%%%%%%%%%%%%%%%%%%%%%%%%%%%%%%%

\subsection{``Magnetogenesis'' in $^3$He-A}

In  various experiments
\cite{Experiment,BigBangNature1,NaturePrimordial} the
flow instability has been
measured using NMR techniques, which means that one needs
an external (real) magnetic field ${\bf H}$. Such a field adds an
additional mass  to the ``hypercharge gauge field''
${\bf A}$ due to the spin-orbital interaction in
Eq.~(\ref{SpinOrbit}).
Even at low $T$ the instability then occurs only above some critical value of
the counterflow velocity $w_0$ (or correspondingly chemical potential of right
electrons $\mu_R$). The critical value $\mu_{R}^{cr}$  depends on  $T$ and $H$
and approaches the value of order $p_F\sqrt{g_D/\rho_s}$ in the limit of
large
$H$.

When this helical instability develops in $^3$He-A, the final result is
the formation of the ${\hat{\bf l}}$-texture which corresponds to the free
energy minimum in the rotating vessel.  This is the periodic ${\hat{\bf
l}}$-texture, whose elementary cell represents the
Anderson-Toulouse-Chechetkin (ATC) continuous vortex in
Fig.~\ref{ContinuousMomentogenesis}. The presence of ATC vortices and their
number is extracted from the NMR absorption spectrum, which contains
the satellite peaks coming from different types of vortices
\cite{PhaseDiagram}. The position of the satellite peak indicates the type of
 vortex, while the intensity is proportional to the number of vortices of
this type.  The satellite peak for the ATC vortices
is shown in Fig.~\ref{experiment}.

%%%%%%%%%%%%%%%%%%%%%%%%%%%%%%%%%%%%%%%%%%%%%%%%%%%%%%%%%%
\begin{figure}[!!!t]
%\centerline{\epsfxsize=0.40\textwidth\epsfbox{Lammi10.eps}}
%\bigskip
\begin{center}
\leavevmode
\epsfig{file=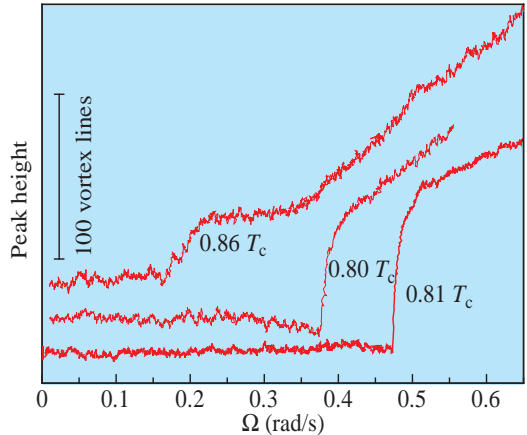,width=0.8\linewidth}
\caption[experiment2]
    {Time dependence of the satellite peak height of the continuous vortices.
Initially vortices are not present in the vessel. When the velocity of the
counterflow
${\bf w}$ in the ${\hat{\bf l}}_0$ direction (corresponding to the chemical
potential
$\mu_R$ of the chiral electrons)   exceeds some critical value, the instability
occurs and the container  becomes filled with the ${\hat{\bf l}}$-texture
(hypermagnetic field) forming the vortex array.}
\label{experiment2}
\end{center}
\end{figure}
%%%%%%%%%%%%%%%%%%%%%%%%%%%%%%%%%%%%%%%%%%%%%%%%%%%%%%%%%%

In the experiment carried out in Helsinki the initial state did not contain
vortices. Then the vessel was put into rotation with some angular
velocity $\Omega$. If the velocity is small enough, one has only
counterflow and no vortex texture. This means that  there is a nonzero
``chemical potential of right electrons'',
$\mu_R=p_F\Omega r$, where $r$ is the distance from the axis
of the rotating vessel,
while the ``hypermagnetic'' field is absent.   Accelerating the vessel further
one finally reaches the critical value $\mu_{R}^{cr}$ at the  wall of
container,
$r=R$, where the counterflow is maximal.  At this moment the instability
occurs,
which is observed by the Helsinki group as  a
jump in the  height of the vortex peak (see  Fig.~\ref{experiment2}). The peak
height jumps from zero to the magnitude corresponding to a
vortex array with nearly the  equilibrium number of vortex lines. This means
that counterflow has been
 essentially removed. The counterflow (which carried the fermionic charge of
matter) has thus been converted to a vortex ${\hat{\bf l}}$-texture
(hypermagnetic field).

The magnitude of $\mu_R^{cr}$ found from experiments
\cite{Experiment}  is in good quantitative agreement with the theoretical
estimation of the mass of the ``hyperphoton'' determined by the spin-orbit
interaction in Eq.~(\ref{SpinOrbit}): $\mu_{R}^{cr} \sim
p_F\sqrt{g_D/\rho_s}$.  Thus the Helsinki experiments model the nucleation of
the hypermagnetic field for different masses of the ``hyperphoton''.
The flow instability in the limit when
the contribution to the ``hyperphoton'' mass from the real external magnetic
field
$H$ is zero has also been investigated: First the field
$H$ was turned off and after the instability had occurred the field was
switched on again and the created ``hypermagnetic field'' was measured. In this
case it was observed that $\mu_R^{cr}$ was significantly reduced.

\section{AXION in $^3$He-A}

We discussed how the  quasiparticles and the ${\hat{\bf l}}$-texture
can exchange the fermionic charge -- the linear momentum -- due to the axial
anomaly. There is yet another phenomenon:  The
${\hat{\bf l}}$-texture and the moving superfluid
vacuum can also exchange momentum. Thus the  ${\hat{\bf l}}$-texture serves
as an intermediate object which allows to transfer the fermionic charge from
the condensate (vacuum) motion to the quasiparticles (matter), as was
discussed in Sec.IIIA. In this sense the ${\hat{\bf l}}$-texture plays the same
role as quantized vortices in superfluids and superconductors. This again
shows the common properties of continuous ${\hat{\bf l}}$-textures (with
continuous vorticity) and quantized singular vortices, which are related to the
gap nodes: In the ${\hat{\bf l}}$-textures the gap nodes are lying
in momentum
space, while in the most symmetric quantized vortices of conventional
superconductors and also in the most symmetric cosmic strings the nodes are in
real space -- in the cores of vortices, where the symmetry is restored and
fermions are massless. The transformation between
the real-space zeroes and the momentum-space zeros \cite{Zeroes1} actually
occurs
when the singular core of the vortex experiences an additional symmetry
breaking, as was observed for the $^3$He-B vortices. The relation of both
types of zeroes to the axial anomaly was discussed in
\cite{Zeroes2}.

Consider now the exchange between the superfluid vacuum and the
texture. The momentum density of the superfluid vacuum along the equilibrium
${\hat{\bf l}}_0$-vector is $m_3n_3v_{sz}$. The momentum exchange follows from
the anomalous nonconservation of the momentum Eq.~(\ref{MomentoProduction2}):
\begin{equation}
  m_3n_3(\partial_t  v_{sz}  +   \partial_z \mu_3)= {p_F\over 2\pi^2}
\left(\partial_t{\bf A}\cdot {\bf\nabla}\times{\bf A}\right)~.
\label{CondensateMomentumNonconservation}
\end{equation}
where $\mu_3$ is the real chemical potential of $^3$He atoms, which also
determines the speed of sound in $^3$He-A: $s^2=n_3 d\mu_3/dn_3$;
${\bf A}=p_F
\delta {\hat{\bf l}}$. In what follows, we
consider only the $z$- and $t$-dependence of all
variables.

Let's introduce a variable $\theta$
which is dual to the potential $\Phi$ of the
superflow:
\begin{equation}
 \partial_t\theta= -{p_F\over 2m_3}\partial_z \Phi=- p_F
v_{sz}~,
\label{AxionField1}
\end{equation}
\begin{equation}
\partial_z\theta= -{p_F\over 2m_3s^2}\partial_t \Phi= {p_F\over
s^2}\delta\mu_3 ~,
\label{AxionField2}
\end{equation}
We can now write down the  Lagrangian whose variation gives rise to the
anomalous nonconservation of the condensate momentum in
Eq.~(\ref{CondensateMomentumNonconservation}):
\begin{equation}
{1\over 2\pi^2} \theta \left(\partial_t\vec A\cdot \vec\nabla\times\vec
A\right) + {n_3m_3\over 2 p_F^2} \left( s^2 (\partial_z\theta)^2-
(\partial_t\theta)^2\right)~.
\label{AxionAction}
\end{equation}
 This is nothing but the action for the axion field $\theta$, which interacts
with the CP violating combination $F^{\mu\nu}\tilde F_{\mu\nu} \propto {\bf
E}\cdot {\bf B}$
\cite{Axions}.  The Joice-Shaposhnikov scenario of the exponential growth
of magnetic field can be also realized if instead of the excess of the right
electrons one has the time dependent axionic field \cite{CarollField}. The role
of the chemical potential is now played by $\partial_t  \theta$. In our case
of superflow this again corresponds to the superfluid velocity according to
Eq.(\ref{AxionField1}).

In $^3$He-A the axion corresponds to  sound waves --
propagating oscillations of two conjugated variables, the phase $\Phi$,
related to rotations of the fundamental
triad, and the particle density $n_3$.  The anomalous
first term in Eq.~(\ref{AxionAction}) can be also obtained from
Eq.~(\ref{Lphi2}). The speed of sound is
$s^2=(1/3)v_F^2(1+F_0)(1+F_1/3)$, where $F_0$ and $F_1$ are Fermi-liquid
parameters, and is the same in superfluid $^3$He-A ($T<T_c$) and in
normal liquid $^3$He
($T>T_c$). In
distinction
from the orbital waves (``electromagnetic waves''), the speed of sound
$s$ is isotropic and does not coincide with any of the two speeds of light,
$c_\perp$ or
$c_\parallel$, though one can expect that the axion propagating along $z$ must
have the parallel speed of ``light'' $c_\parallel=v_F$.  What is the reason?
It is a property of the superfluid $^3$He-A:

Both modes, ``photon'' (orbital wave) and ``axion'' (sound wave) are collective
bosonic excitations of the fermionic system and are obtained by integration
over the fermions. In the case of ``photons'' the relevant
region of the  integration over the fermions is concentrated close to the gap
nodes due to the logarithmic divergency. Near the nodes the fermions are
relativistic and are described by the Lorentzian metric $g^{\mu\nu}$.
It follows that
 the effective ``photons'' are described by the same metric and therefore
the speed of light is the same as the speed of the massless fermion propagating
in the same direction.
On the other hand, the relevant   region of the  integration, which is
responsible for the  spectrum of the ``axion'' mode, is  far from the gap
nodes. Consequently the axion spectrum  does not even depend on the
existence of
the gap nodes and induces its own effective metric.

\section{DISCUSSION}

In principle one can introduce a model system with favourable
parameters,
such that for all collective modes the integration over the fermions is
concentrated mostly in the region where the fermions are Lorentzian.  In this
case the low energy dynamics of photons, axions, gravitons, etc., will be
determined by the same  Lorentzian metric as that of the fermions.
In the low-energy
corner,  one then obtains the effective relativistic quantum field theory and
effective quantum gravity with the same speed of light for all bosons and
fermions.

It is quite possible that in this ideal case the cosmological constant
vanishes. This follows from the fact that
Eq.~(\ref{E^2relativistic}) for the spectrum of massless quasiparticles can be
multiplied by an arbitrary scaling factor $a^2$,
which does not change the energy spectrum,
but changes the contravariant metric tensor: $g^{\mu\nu} \rightarrow a^2
g^{\mu\nu}$. Since physics cannot depend on such formal conformal
transformation, the effective low-energy Lagrangian for gravity cannot
depend on $a^2$ and thus the cosmological term $\int d^3xdt ~\Lambda\sqrt{-g}$
is prohibited (a discussion of the role of the scale invariance
for vanishing cosmological constant is found in Ref.\cite{AdlerCosmConstant}).

This situation is somewhat similar to that which occurs in the normal
Fermi-liquid where the role of the parameter $a^{-1}$ is played by the
quasiparticle spectral weight $Z$ -- the residue of the  Green's function
at the
quasiparticle pole. The low-energy properties of this system, described by the
Landau phenomenological Fermi-liquid theory, do not depend on
$Z$. The Landau Fermi-liquid differs from our system only in
the topology of the spectrum of the low-lying fermionic excitations: The
Fermi-surface instead of the Fermi-points -- the gap nodes, is present there.

Note that the Fermi-surface and the point node are the only topologically
stable features of the fermionic spectrum. They are described by
$\pi_1$ and $\pi_3$ topological invariants respectively and thus are robust to
any modification of the system. These two classes exhaust the topologically
stable gapless Fermi systems. In Landau theory, which deals with the
Fermi-surface class of Fermi liquids, the low-energy bosonic collective modes
are related to the dynamical deformations of the Fermi surface. In the
point-node class  of Fermi liquids, the corresponding collective motion comes
from the dynamics of the nodes. This dynamics gives rise to effective gravity
and  effective electromagnetic fields.

The fundamental constants  in these effective theories are
determined by the position of the node, $p_F$, and by the slopes of the energy
$E$ of the quasiparticle as a function of its momentum
${\bf p}$ at the node. There are three such parameters in $^3$He-A: the  Fermi
velocity $v_F$, the Fermi momentum $p_F$ and the gap amplitude $\Delta_0$.
They give the parallel speed of light $c_\parallel=v_F$; the transverse speed
of light  $c_\perp=\Delta_0/ p_F$;  the Planck energy $\Delta_0$; the running
coupling constant in Eq.~(\ref{RunningCoupling}); the masses of the hyperphoton
in Eqs.~(\ref{Mass1},\ref{Mass2rel});  gravitational constant $G\sim
 \Delta_0^{-2}$
\cite{GravitationalConstant} and cosmological constant
$\sim \Delta_0^4$ \cite{Volovik1986}; etc.  Since all 3 initial parameters are
in principle temperature dependent, the fundamental constants are not constants
in the effective theories. For example the speed of light depends on
temperature and also on the photon energy: $\delta c/c \sim
(E/E_{\rm Planck})^2$. The larger (linear) effect, $\delta c/c \sim
 E/E_{\rm Planck} $, was discussed in \cite{Amelino}.

We discussed only 3 experiments in superfluid $^3$He-A related  to the
properties of the electroweak vacuum. In all of them the chiral
anomaly is an important mechanism.  It regulates the nucleation of the
fermionic charge from the vacuum, as observed in
Manchester \cite{BevanNature}, and the inverse process of the nucleation of the
effective magnetic field from the fermion current, as observed in
Helsinki \cite{Experiment,NaturePrimordial}.

There are many other connections
between superfluid $^3$He and different branches of physics which should be
explored. For example, we can simulate  phenomena related to the effective
gravity, such as the cosmological constant, quantum properties of the event
horizon, vacuum instability in strong gravity, torsion strings and even
inflation.  In principle a nonequilibrium vacuum state can be constructed in
which the speed of light
$c_\perp=\Delta_0/p_F$ decreases exponentially with time. In cosmological
language this implies inflation, since the length scale in the spatial
metric $g_{ik}$ is growing exponentially. This would  allow for a
study of the
development of perturbations during inflation.

Till now we considered the properites related to one pair of nodes only. If one
takes into account that in $^3$He-A there is a two-fold degeneracy related to
two spin projection of the $^3$He atom, the number of
fermiomic and bosonic degrees of freedom increases. It appears that with these
new degrees of freedom, the system transforms to a
$SU(2)$ gauge theory: The conventional spin degrees of freedom of $^3$He atoms
form the $SU(2)$ isospin, while some collective modes of the order parameter
(the spin-orbital waves) behave as $SU(2)$ gauge bosons \cite{exotic}.

There are several ways of extending the model, in which higher local
and global symmetry groups can naturally arise. (1) One can imagine an initial
normal state of condensed matter consisting of $n=3,4,$ etc. degenerate sheets
of the Fermi-surface. Then the superconducting/superfluid  Cooper pairing will
lead to
$n$-fold degeneracy of gap nodes, which in turn gives rise to the effective
local
$SU(n)$  group in the low-energy corner. (2) The number of gap nodes on
each Fermi-surface can be also larger than 2. For example the so called
$\alpha$-state of $^3$He \cite{VollhardtWolfle} contains 8 gap nodes per
Fermi-surface and thus 8 elementary relativistic fermions in the vicinity of
the nodes. The fluctuations of positions of these nodes are equivalent to
several gauge fields. In high-T$_c$ superconductivity each Fermi-sheet
(actually the Fermi-circle since this kind of
superconductivity effectively occurs in the two-dimensional
space of the CuO plane) contains 4 gap nodes. The
corresponding Weyl-like Hamiltonian for 4 fermions and the corresponding gauge
fields have been discussed for this material in
Ref.\cite{Nersesyan}. Thus in principle it appears possible to
construct a model which has as
many fermionic and bosonic degrees of freedom as needed in Grand
Unified Theories, including the different
generations of fermions. Of course, the construction of a suitable
condensed matter system
corresponding to such a model is not a simple undertaking.

\section{Acknowledgements}

I thank Uwe Fischer, Henry Hall, John Hook, Ted Jacobson, Michael Joyce, Tom
Kibble, Matti Krusius, Robert Laughlin, Misha Shaposhnikov,  Alexei
Starobinsky, Tanmay Vachaspati and Shuqian Ying for fruitful discussions.

\vfill\eject


\begin{references}

\bibitem{Wilczek}  F. Wilczek, ``The Future of Particle Physics as a Natural
Science'', hep-ph/9702371.

\bibitem{Nielsen} H.B. Nielsen, ``Field theories without fundamental (gauge)
symmetries'', preprint NBI-HE-83-41.

\bibitem{Sakharov} A.D. Sakharov, Sov. Phys. Doklady {\bf 12}, 1040 (1968).

\bibitem{Zeldovich} Ya.B. Zeldovich, ``Interpretation of electrodynamics as a
consequence of quantum theory'',   Pis'ma ZhETF {\bf 6}, 922 (1967) [JETP
Lett. {\bf 6} 345 (1967)].


\bibitem{Laughlin} R. Laughlin, ``Gauge Theories from Nothing in Condensed
Matter: Are Quantum Critical Points Supersymmetric Field Theories'',  talk
at the Symposium on QUANTUM PHENOMENA at LOW TEMPERATURES, Lammi, 7-11
January, 1998; see also  ``Parallels Between Quantum Antiferromagnetism and the
Strong Interactions'', Proceedings of the Inauguration Conference of the
Asia-Pacific Center for Theoretical Physics, Seoul National University, Korea
4-10 June 1996, ed. by Y. M. Cho, J. B. Hong, and C. N. Yang (World Sci.,
Singapore,1998),  cond-mat/9802180.


\bibitem{VolovikGravity1997} G.E. Volovik, ``Simulation of Quantum Field Theory
and Gravity in Superfluid He-3'', Low Temp. Phys. (Kharkov) {\bf 24}, 127
(1998), cond-mat/9706172.

\bibitem{JacobsonVolovik} T. Jacobson and G.E. Volovik , ``Event
horizons and ergoregions in $^3$He'', cond-mat/9801308, to appear in  Phys.
Rev.
{\bf D~ 58}  (1998); G.E. Volovik, ``Induced Gravity in Superfluid 3He'',
cond-mat/9806010.

\bibitem{RotatingCore}  N. B. Kopnin and G. E. Volovik,  ``Rotating
vortex core:
An instrument for detecting the core excitations'',
 Phys. Rev. {\bf B~57} 8526 (1998).

\bibitem{UnruhSonic} W.G. Unruh,   ``Experimental black-hole evaporation?'',
Phys. Rev. Lett. {\bf 46} 1351 (1981);  ``Sonic analogue of black holes
and the effects of high frequencies on black hole evaporation, ''Phys. Rev.
{\bf
D~ 51} 2827 (1995).

\bibitem{Jacobson1991} T. Jacobson,  ``Black hole evaporation and ultrashort
distances'', Phys. Rev. {\bf D~ 44} 1731 (1991).

\bibitem{Visser1997} M. Visser, ``Acoustic black holes: horizons, ergospheres,
and Hawking radiation'', Class. Quant. Grav. {\bf 15}, 1767 (1998).

\bibitem{exotic} G.E. Volovik,  Exotic properties of superfluid
$^3$He, World Scientific, Singapore-New Jersey-London-Hong Kong,
1992.

\bibitem{VolovikVachaspati}  G.E. Volovik and T. Vachaspati, ``Aspects of
$^3$He and the standard electroweak model'', Int. J. Mod. Phys. {\bf B~10},
471 (1996).

\bibitem{SimonLee} S.H. Simon and P.A. Lee, ``Scaling of the quasiparticle
spectrum for $d$-wave superconductors'', Phys. Rev. Lett., {\bf 78}, 1548
(1997); {\bf 78}, 5029 (1997).

\bibitem{WenEdgeStates} X.G. Wen, Phys. Rev. B, {\bf 43}, 11025 (1991).

\bibitem{StoneEdgeStates} M. Stone, Annals of Phys., {\bf 207}, 38 (1991).

\bibitem{LaughlinT-symmetry} R.B. Laughlin,  ``Magnetic Induction of d + i d
Order in High-Tc Superconductors'', cond-mat/9709004.

\bibitem{VolovikT-symmetry} G.E. Volovik, ``On edge states in superconductor
with time inversion symmetry breaking'',   JETP Lett. {\bf 66 },  522
(1997).

\bibitem{ColorSuperconductivity1}  S. Ying, ``The
Quantum Aspects of Relativistic
Fermion Systems with Particle Condensation'', hep-th/9711167;  ``On the Local
Finite Density Relativistic Quantum Field Theories'',
hep-th/9802044.
\bibitem{ColorSuperconductivity2}  M. Alford, K. Rajagopal, F. Wilczek,
``QCD at
Finite Baryon Density: Nucleon Droplets and Color Superconductivity'',
hep-ph/9711395.
\bibitem{ColorSuperconductivity3} F. Wilczek, `` From Notes to Chords in QCD'',
hep-ph/9806395.

\bibitem{Kajantie}  K. Kajantie, M. Laine, K. Rummukainen, and M.
Shaposhnikov,  ``Is there a hot electroweak transition at $m_H\geq
m_W$?'',  Phys. Rev. Lett. {\bf 77}, 2887 (1996), see also discussion in  A.
Rajantie, ``Vortices and the Ginzburg-Landau phase transition'', Proceedings
of the Symposium on QUANTUM PHENOMENA at LOW TEMPERATURES, Lammi, 7-11 January,
1998 (this volume).

\bibitem{Adler1969} S. Adler, ``Axial-vector vertex in spinor
electrodynamics'', Phys. Rev. {\bf 177}, 2426  (1969).

\bibitem{BellJackiw1969} J.S.Bell and R.Jackiw, ``A PCAC
puzzle:$\pi^0\rightarrow \gamma\gamma$ in the $\sigma$-model'', Nuovo Cim.
{\bf
A60}, 47--61 (1969).

\bibitem{Chechetkin} V.R. Chechetkin, JETP,  {\bf 44}, 706 (1976).

\bibitem{AT} P.W. Anderson and G. Toulouse, Phys. Rev. Lett.,
{\bf 38}, 508 (1977).

\bibitem{Volovik1992} G.E. Volovik, ``Hydrodynamic action for   orbital and
superfluid dynamics of $^3$He-A at $T=0$'',   JETP {\bf 75}, 990 (1992).

\bibitem{AB} G.E. Volovik,   ``Three nondissipative forces on a  moving vortex
line in superfluids and superconductors'', JETP Lett. {\bf  62}, 65
(1995); ``Vortex vs spinning  string:  Iordanskii
force and gravitational Aharonov-Bohm effect''.

\bibitem{Kopnin1993} N.B. Kopnin, ``Mutual friction in
superfluid  $^3$He. II. Continuous vortices in $^3$He-A at low temperatures'',
Phys. Rev.   {\bf B~47}, 14354 (1993).

\bibitem{BevanNature}  T.D.C. Bevan, A.J. Manninen, J.B. Cook, J.R. Hook,
H.E. Hall, T. Vachaspati and G.E. Volovik,
``Momentogenesis by $^3$He vortices: an
experimental  analogue of primordial baryogenesis'',  Nature,  {\bf 386},
689  (1997).

\bibitem{BevanJLTP}  T.D.C. Bevan, A.J. Manninen, J.B. Cook, H. Alles, J.R.
Hook, H.E. Hall,
``Vortex mutual friction in superfluid $^3$He vortices'', J. Low Temp.
Phys. {\bf
109},423  (1997).

\bibitem{Dolgov} A.D. Dolgov, ``Non-GUT baryogenesis'', Rep. Prog. Phys.,
{\bf 222}, 309 (1992).

\bibitem{Turok} N.Turok, in ``Formation and interaction of
topological defects'', eds.  A.C.Davis, R.Brandenberger
(Plenum Press, New York,London),1995, pp. 283--301.

\bibitem{ewitten} E. Witten, Nucl. Phys.  {\bf B249},
557 (1985).


\bibitem{tvgf} T. Vachaspati and G.B. Field, ``Electroweak
string configurations with baryon number'', Phys. Rev. Lett. {\bf  73},
373--376 (1994);  {\bf 74}, 1258(E) (1995).

\bibitem{jgtv} J. Garriga and T. Vachaspati, ``Zero modes on
linked strings'', Nucl. Phys. {\bf B~438}, 161  (1995).


\bibitem{barriola}  M. Barriola,  ``Electroweak strings produce
baryons'', Phys. Rev. {\bf D~51}, 300 (1995).

\bibitem{gstv}  G.D. Starkman and T. Vachaspati , ``Galactic
cosmic strings as sources of primary antiprotons'', Phys. Rev. {\bf D~53},
6711  (1996).

\bibitem{KopninVolovikParts} N.B. Kopnin, G.E. Volovik, and \"U. Parts
 ``Spectral flow in vortex dynamics of 3He-B and superconductors'',
Europhys. Lett. {\bf 32} 651 (1995).

\bibitem{StoneSpectralFlow} M. Stone, ``Spectral flow, Magnus force and mutual
friction via the geometric optics limit of Andreev reflection'', Phys. Rev.
{\bf B~54}, 13222 (1996).

\bibitem{JoyceShaposhnikov}  M. Joyce, M. Shaposhnikov, ``Primordial
magnetic fields, right electrons, and the abelian anomaly'', Phys.
Rev. Lett., {\bf 79}, 1193  (1997).

\bibitem{GiovanniniShaposhnikov} M. Giovannini and E.M.
Shaposhnikov, ``Primordial
hypermagnetic fields and triangle anomaly'',  Phys.Rev. {\bf D~57} 2186 (1998).

\bibitem{NaturePrimordial}  M. Krusius,   T. Vachaspati
 and G.E. Volovik
``Flow instability in 3He-A as analog of generation of hypermagnetic
field in early Universe'', cond-mat/9802005.

\bibitem{Experiment}  V.M.H. Ruutu, J. Kopu, M. Krusius, U. Parts, B.
Pla¡ais, E.V. Thuneberg, and W. Xu , ``Critical velocity of
vortex nucleation in rotating superfluid 3He-A'',
Phys. Rev. Lett.,{\bf 79}, 5058 (1997).

\bibitem{VollhardtWolfle} D. Vollhardt, P. and P. W\"olfle,  ``The
superfluid phases of helium 3'',  Taylor and Francis, London - New York -
Philadelphia, 1990.

\bibitem{OrbitalMomentum} G.E. Volovik,  ``Orbital momentum of vortices and
textures   due to spectral flow through the gap nodes:  Example of the 3He-A
continuous vortex'', JETP Lett. {\bf  61}, 958  (1995).

\bibitem{Volovik1986} G.E. Volovik,   ``Analog of gravity in superfluid
   $^3He-A$'', JETP Lett. {\bf 44}, 498  (1986).

\bibitem{Jegerlehner} F. Jegerlehner, "The ``ether-world´´ and elementary
particles", hep-th/9803021.

\bibitem{Weldon} H.A. Weldon, ``Covariant calculations at finite temperature:
The relativistic plasma'', Phys. Rev. {\bf D~26}, 1394 (1982).

\bibitem{Wolfle} P. W\"olfle, in ``Quantum Statistics and the Many Body
Problem'' (S.B. Trickey, W.P. Kirk, and J.W. Duffy, Eds.) p.9. Plenum, New
York, 1975.

\bibitem{Volovik1975}  G.E. Volovik, ``Orbital momentum  and orbital waves in
the anisotropic A-phase of superfluid $^3He$'', JETP Lett. {\bf 22}, 108
(1975);  ``Dispersion of the orbital waves in the A-phase of superfluid
$^3$He'', JETP Lett. {\bf 22}, 198   (1975).

\bibitem{LeggettTakagi} A.J. Leggett and S. Takagi, ``Orientational Dynamics of
Superfluid $^3$He: A ``Two-Fluid'' Model. II. Orbital Dynamics'',  Ann. Phys.
{\bf 110}, 353  (1978).

\bibitem{DittrichGies} W. Dittrich, and H. Gies, ``Light propagation in
non-trivial QED vacua'',  hep-ph/9804375.

\bibitem{GravitationalConstant} G.E. Volovik, ``Gravity of monopole and string
and gravitational constant  in $^3$He-A'',   Pis'ma ZhETF {\bf
67}, 666 (1998), cond-mat/9804078.

\bibitem{ThesisVollhardt} D. Vollhardt, ``Stability of supeflow and
related textural transformations in superfluid $^3$He'', PhD
Thesis, Hamburg, 1979.

\bibitem{BigBangNature1} V.M.H.Ruutu, V.B.Eltsov, A.J.Gill,
T.W.B Kibble, M.Krusius, Yu.G.Makhlin, B.Placais,
G.E.Volovik and Wen Xu, ``Vortex formation in neutron irradiated
$^3$He as an analogue of cosmological defect formation'',
Nature {\bf 382}, 334   (1996).

\bibitem{PhaseDiagram} \"U. Parts, J.M. Karim\"aki, J.H.
Koivuniemi,  M. Krusius, V.M.H. Ruutu, E.V. Thuneberg, and
G.E. Volovik, ``Phase diagram of vortices in superfluid $^3$He-A'',
Phys. Rev. Lett. {\bf 75}, 3320  (1995).

\bibitem{Zeroes1} G.E. Volovik, V.P. Mineev,   ``Current in
  superfluid Fermi liquids and the vortex core structure'', JETP {\bf 56},
  579   (1982)

\bibitem{Zeroes2} G.E. Volovik,  ``Action for anomaly in fermi superfluids:
  quantized vortices and gap nodes'', JETP {\bf  77},   435   (1993).

\bibitem{Axions} M.S. Turner, ``Windows on the axion'', Phys. Rep. {\bf
197}, 67
(1990);  J.E. Kim, ``Cosmic Axion'', astro-ph/9802061.

\bibitem{CarollField} M.S. Turner, and L.M. Widrow,
Phys. Rev. {\bf D~37}, 2743 (1988); S.M. Carroll, and  G.B. Field, ``Primordial
Magnetic Fields that Last?'', astro-ph/9807159.

\bibitem{AdlerCosmConstant} S.L. Adler, ``A strategy for a vanishing
cosmological constant in the presence of scale invariance
breaking'', Gen. Rel. Grav. {\bf 29} 1357  (1997).

\bibitem{Amelino} G. Amelino-Camelia, J. Ellis, N.E. Mavromatos, D.V.
Nanopoulos, and S. Sarkar, ``Test of quantum gravity from observations of
$\gamma$-ray bursts'', Nature,  {\bf 393}, 763 (1998).

\bibitem{Nersesyan} A.A. Nersesyan, A.M. Tsvelik and F. Wenger,  ``Disorder
effects in two-dimensional $d$-wave superconductors'', Phys. Rev. Lett.,  {\bf
72}, 2628 (1994).



\end{references}
\end{document}